\documentclass[graybox, envcountchap]{svmult}

\usepackage{mathptmx}        
\usepackage{amsmath}
\usepackage{amssymb}
\usepackage{color}
\usepackage{helvet}          
\usepackage{courier}         
\usepackage{dirtree}

\usepackage{makeidx}        
\usepackage{graphicx}        
\usepackage{subfig}

\usepackage{multicol}        
\usepackage[bottom]{footmisc}

\usepackage{hyperref}        
\hypersetup{colorlinks=true,urlcolor=blue}

\usepackage[misc]{ifsym}

\makeindex             

\begin{document}

\def\be{\begin{equation}}
\def\ee{\end{equation}}
\def\bea{\begin{eqnarray}}
\def\eea{\end{eqnarray}}
\newcommand{\tb}[1]{\textbf{\texttt{#1}}}
\newcommand{\actaa}{Acta Astronomica}
\newcommand{\laa}{\mathcal{L}}
\newcommand{\ba}{\mathcal{B}}
\newcommand{\Sie}{\mathcal{S}}
\newcommand{\Mie}{\mathcal{M}}
\newcommand{\La}{\mathcal{L}}
\newcommand{\Em}{\mathcal{E}}
\newcommand{\Pa}{\mathcal{P}}
\newcommand{\mso}{\mathrm{mso}}
\newcommand{\mbo}{\mathrm{mbo}}
\definecolor{LinkBlue}{RGB}{6,69,173}
\definecolor{DarkBlue}{RGB}{11,0,128}
\definecolor{red}{rgb}{1,0.,0.}
\newcommand{\rtb}[1]{\textcolor[rgb]{1.00,0.00,0.00}{\tb{#1}}}
\newcommand{\gtb}[1]{\textcolor[rgb]{0.17,0.72,0.40}{\tb{#1}}}
\newcommand{\ptb}[1]{\textcolor[rgb]{0.77,0.04,0.95}{\tb{#1}}}
\newcommand{\btb}[1]{\textcolor[rgb]{0.00,0.00,1.00}{\textbf{#1}}}
\newcommand{\otb}[1]{\textcolor[rgb]{1.00,0.50,0.25}{\tb{#1}}}
\newcommand{\non}[1]{{\LARGE{\not}}{#1}}

\newcommand{\cc}{\mathrm{C}}

\newcommand{\il}{~}
\newcommand{\la}{\mathcal{A}}
  \newcommand{\Qa}{\mathcal{Q}}
\newcommand{\Sa}{\mathcal{\mathbf{S}}}
\newcommand{\Ta}{{\mbox{\scriptsize  \textbf{\textsf{T}}}}}
\newcommand{\Ca}{\mathcal{\mathbf{C}}}
  \newcommand{\MB}{\textbf{MB}}
\renewcommand{\hbar}{\mathchar'26\mkern-9mu h}

\title{Naked singularities  and black hole Killing horizons}
\author{Daniela Pugliese and Hernando Quevedo}
\institute{Daniela Pugliese (\Letter) \at Research Centre for Theoretical Physics and Astrophysics, Institute of Physics,
  Silesian University in Opava,
 Bezru\v{c}ovo n\'{a}m\v{e}st\'{i} 13, CZ-74601 Opava, Czech Republic, \email{daniela.pugliese@physics.slu.cz}
\and Hernando Quevedo \at Instituto de Ciencias Nucleares, Universidad Nacional Aut\'onoma de M\'exico, AP 70543, Ciudad de M\'exico 04510, MEXICO  \email{quevedo@nucleares.unam.mx}}
%
%
\maketitle

\abstract{In this chapter, we study  special photon orbits defined by means of Killing vectors and present a framework based on the properties of such null orbits. For concreteness, we restrict ourselves to the case of axially symmetric spacetimes describing either black holes  with Killing horizons or naked singularities. The null-orbits framework is then applied to analyze properties of naked singularities and concepts of black hole thermodynamics. }


%
%
\section{Introduction}
The creation of  a  naked singularity  (\textbf{NS})  through   accretion onto   a (extreme) black hole \textbf{(BH)} is    forbidden.
In particular, the horizon area cannot decrease
 in any process with suitable conditions on the energy and  no repulsive gravity. More precisely,   the total
horizon area of (classical) \textbf{BHs} cannot decrease over time (Hawking's area theorem -\cite{Hawking71,Hawking74,Hawking75}). Observational data of  GW150914 have been recently  considered  as an  observational confirmation of Hawking's black-hole area theorem-- \cite{Isi:2020tac}.
 However,  assuming    the existence of \textbf{NS}s,  the reverse process, consisting in the conversion, at some point of  the singularity formation or   evolution, into a \textbf{BH}   is still debated in different contexts.
 A classical method for extracting energy from a spinning  \textbf{BH} and  from a hypothetical \textbf{NS}  is,  for example, the
Penrose  process, which is based on  the  frame-dragging effects of  the ergoregion of a spinning \textbf{BH} --\cite{Penrose69}.
 In this region, a small body  may have   negative energy, as measured by an observer at infinity and, if
fragmented into two parts, and    the part with negative-energy  will be  swallowed by the \textbf{BH}, it will decrease its (total) mass, while
the other part,  with positive energy, can be   ejected. In this process,  the  Kerr \textbf{BH} rotational  energy   decreases and consequently, the  Kerr \textbf{BH}  area  (determined by its  outer horizon)  increases.

In this chapter, we focus on \textbf{BH}  spacetimes  with  Killing  horizons and  on   naked singularity solutions of the corresponding field equations.
It is possible to study the singularities  using   light surfaces related to the  background Killing fields, and in particular to the   generators of Killing horizons. Special null orbits associated to the Killing vectors of the geometry   can generate a useful framework for the study  of the more diverse   geometries, such as wormholes, accelerating black holes,  or binary  black hole systems\footnote{See \cite{wormhole} for the study of the light surfaces in    static and spinning   wormhole solutions,   black holes  immersed in  external (perfect fluid) dark matter,   spacetimes with  (Taub) NUT charge, acceleration,  magnetic charge, and cosmological constant, binary Reissner--Nordstr\"om   black holes,   a solution of  a  (low--energy effective)  heterotic string theory, and  the  $(1+2)$ dimensional BTZ geometry, where there are
 cosmological  and acceleration horizons, or  internal solutions matching  exact vacuum solutions of Einstein equations. Properties of  the \textbf{BH} thermodynamics for a Mini-Super-Space, semi-classical polymeric \textbf{BH} have  been also  discussed  in \cite{LQG} in this framework.}, pointing  out the existence of common properties of the light surfaces in different  spacetimes.

Here, we will use the properties of the Killing vectors, generators of the \textbf{BH} Killing horizons, limiting ourselves to the cases  of the Kerr and   Kerr--Newman spacetimes.

The Kerr-Newman (\textbf{KN}) solution is an electrically charged, electro--vacuum, asymptotically flat, axially symmetric (and stationary) solution of the Einstein--Maxwell equations, used to describe  the spacetime  around an electrically charged gravitational source, spinning along it symmetry axis. The Kerr solution can be read as a limit of the \textbf{KN} metric when  the electric charge is null.  According to  the "no--hair" theorems,  the  Einstein--Maxwell  \textbf{BH}  solutions are characterized only by  mass, electric charge, and angular momentum; therefore  the Kerr--Newman  spacetime is the   most general
asymptotically flat stationary \textbf{BH} solution of  the Einstein--Maxwell equations.   {On the other hand, there are other stationary axisymmetric solutions of the Einstein--Maxwell equations with additional parameters, such as acceleration, magnetic charge, NUT charge, and cosmological constant, having therefore up to seven parameters. }
	Furthermore, the theorem has been proved  to be violated, for example, by  the   self--gravitating Yang--Mills solitons  or  dilaton fields with   asymptotically--defined global charges.
	The \textbf{KN} spacetime has also as a limiting case
	the static and electrically charged Reissner--Nordstr\"om solution.
	The Schwarzschild  spacetime is a static, neutral and spherically symmetric solution\footnote{In General Relativity, any spherically symmetric solution in  vacuum  must be static (and asymptotically flat), implying that the  { vacuum exterior} solution of a  spherically symmetric gravitating source must be  the  Schwarzschild metric, which is the content of  the  well known Birkhoff's theorem \cite{Birkhoff}. Anticipating what is known  today as \textbf{BH} uniqueness, the inverse of this formulation was  the  first formulation (grounded on  different conceptual bases) of   Israel's  (no--hair) theorem \cite{Israel1,Israel2}.
However, Birkhoff's theorem  has been  generalized in different formulations. In particular, any spherically symmetric and asymptotically flat solution of the Einstein--Maxwell field equations, is static, picturing  the case of the Reissner--Nordstr\"om electro--vacuum solution. (On the other hand, a not asymptotically flat  Einstein--Maxwell  spherically symmetric   solution,  is for example the   Bertotti-Robinson universe\cite{Riegert}).
A further consequence of the  Birkhoff theorem is that  a  (charged)
spherical (uniform) shell of dust  has   flat spacetime   inside
the shell.
If a star is not static but it is spherically symmetric,
the solution outside must be  the Schwarzschild solution if  electrically neutral.  Spherical symmetry
implies stationarity.
(In other words, the Schwarzschild metric  can be derived  without imposing  the metric stationarity).
Furthermore, if  the central source
has a (radial) time evolution  (radially pulsating, spherical collapse, etc),
the exterior metric is the  Schwarzschild (static) spacetime.
Hence,    there are no  gravitational waves
emitted from a spherically symmetric system. (A similar symmetry prevents a spherically symmetric distribution
of charges or  currents  to radiate.
 No
monopole electromagnetic radiation is possible, the  Coulomb solution is  the  only spherically symmetric solution of Maxwell's equations in
vacuum).
On the other hand, it has been  also shown that  static \textbf{BHs}  are not necessarily spherically symmetric or  axially symmetric and viceversa  that non-rotating \textbf{BHs}  are not necessary static
	\cite{Kunz2,Wi2,Puls,Puls1,wormhole}.}.

The \textbf{KN}  \textbf{BH}  horizons are Killing horizons.
A Killing horizon is a hypersurface where a Killing vector of the metric becomes null. Or, more precisely, a Killing horizon  is a null hypersurface,
whose null tangent vector can be normalized to coincide with a Killing vector field.
In other words, a Killing horizon is a  null surface, whose normal is a Killing vector field. Null geodesics,   whose tangent vectors
are normal to a null hypersurface $N$ are called  generators of $N$\footnote{A  portion of a null hypersurface can be also   the Killing horizon of two or more independent Killing vectors--\cite{Mars}.}.
 In particular,
the event horizon of a stationary and  asymptotically flat \textbf{BH} geometry is a Killing horizon (Hawking  rigidity theorem)
 with respect to  a Killing vector, say  $\mathcal{L}^{a}$.
We  shall consider the null vector  $\mathcal{L}^{a}$ in \emph{all} points of the  spacetime, including the case of naked singularities.
The light surfaces considered here are   null  hypersurfaces  defined by   Killing vectors and characterized by a constant parameter  $\omega$, which is the  photon  orbital  frequency.

Within  this framework, we can explore properties enfolding along
the entire family of  geometries, and  the notion of
Killing horizons are reconsidered from a special perspective, where  \textbf{NS} solutions are related to \textbf{BH} solutions.

The relevance  of this new representation, together with the possible conceptual significance, lies  also   in the fact that
 the  entire
family of  solutions  is considered    as a  \emph{unique geometric object}.
In this new framework, we can find   properties  of   the spacetime geometries where several   geometric {quantities}, such  as the \textbf{BH} horizons,  acquire a completely different significance, when considered for the entire family, where naked singularity solutions, for example,  are related to the \textbf{BH}  horizons.

A key element  in the definition of this frame is, therefore, the  \textbf{BH} Killing  horizons.
Horizons define  \textbf{BHs}, fixing their geometrical, thermodynamical and  even quantum properties. Bekenstein  first  suggested the  idea that the  event horizon defines the \textbf{BH}  entropy--\cite{Bekenstein73,Bekenstein75}.
The \textbf{BH}   thermodynamic properties
acquire,  therefore, a  purely  geometric meaning.
 A  \textbf{BH} macrostate is defined and determined only by the mass  $M$, spin $J$,
and electric charge $Q$, while the number of \textbf{BH} microstates increases with the horizon area, which is a
 function of the outer horizon. Hence, the \textbf{BH}  state  may  correspond  to a large number of microstates, leading
 to a very high \textbf{BH} entropy.

 In a very multifaceted sense, 	\textbf{BH} horizons define also its quantum properties. The progenitor  collapse into a \textbf{BH} appears as an  irreversible process of
information degeneration, where  the information   becomes inaccessible  to the distant observers and the \textbf{BH} has no observable
hair.
However, if   \textbf{BHs} have an entropy and a  temperature, this would imply the possibility of \textbf{BH}  radiation. The Hawking  radiation (tightly connected to the Unruh effect) is  explained by effects of quantum fluctuation  of matter fields  in the region close  to the  horizon, which  possibly  could be observed in the  thermal profile.
The derivation of Hawking radiation is,  however, semi-classical   and yet  does not determine  if the \textbf{BH}
can  evaporate completely or   some  remnants are left (in this case, the  emitted radiation may be  entangled with the \textbf{BH}  remnant, which should have entropy). This fact  is at basis of the so called  \textbf{BH}  information paradox.
A  possibility is that the final radiation of the complete evaporation does not ``bring" information of the
 initial matter state (see, for example, \cite{JP16} for a review on the  \textbf{BH} information problem).

 However, a \textbf{BH} singularity is     an  ``active"  gravitational    object in the sense that it can interact with its environment constituted  by matter and fields. Then, a  \textbf{BH}  transition from a state to another, corresponding to a change of the \textbf{BH}
characteristic parameters,  is regulated by the  laws of \textbf{BH}  thermodynamics. Such transition may follow, for example, by a  \textbf{BH}  energy extraction,  which can be detectable by observing   jet   emissions.
 Therefore, we will explore in this framework also the concepts of \textbf{BH} thermodynamics.

The  light surfaces considered here  define particular structures known  as metric Killing  bundles (or more simply metric bundles--\MB s),  enlightening  some properties of  the spacetime causal structure as  spanning in different geometries.
Metric bundles collect all the geometries of the same metric family having a particular characteristic in common. In the  Kerr spacetime, for example, metric bundles collect all geometries having equal photon circular orbiting  frequency. One advantage of the  \MB s   is the establishment of   a new framework of analysis where  the   entire
family of  metric solutions      is studied  as a  unique  geometric object, and
 the  single  spacetime  is a   part of the plane (\emph{extended plane}), where \MB s are defined as curves tangent to the Killing horizons curve, which is the curve in the extended plane representing all the \textbf{BH} horizons. In this way, the    properties of the spacetime solutions are studied  as  unfolding across the spacetimes of the plane.
Different geometries  are, therefore, related  through metric bundles. The extended plane
 establishes also a \textbf{BH}--\textbf{NS} connection,  through the tangency condition, providing a global frame,  in particular, for the study of \textbf{NS} solutions \cite{bundle-EPJC-complete,LQG,observers,ella-correlation,remnants}.
At the base of this relation, for example, in the Kerr spacetime, is the fundamental property that
 all the  \MB s photon circular  orbits are the  frequency of an  inner or outer   Kerr \textbf{BH} horizon. These photon circular orbits are known as horizon \emph{replicas}. A  consequence of this fact is that the horizons curves emerge as the envelope surfaces of all the curves associated to the \MB s in the extended plane.


\medskip

In details, the structure of this chapter is as follows:
in Sec.\il(\ref{Sec:axila}), we introduce
  the Kerr-Newman geometry.
In Sec.\il(\ref{Sec:LS-MBS}), we build   the \MB s framework using the properties of special light surfaces and the   \textbf{BH}  horizons, discussing some of their characteristics and relevance for  \textbf{BH} and \textbf{NS} spacetimes, while in the following  sections we  detail   these definitions for the Kerr geometry.
We start in Sec.\il(\ref{Sec:majo-sure})  examining  the Killing horizon definitions in the context of metric bundles.
Stationary observers and metric bundles are the focus of Sec.\il(\ref{Sec:Station-ob}).
Light surfaces  are investigated in Sec.\il(\ref{Sec:light-surfaces-BottleN}).
\MB s characteristics in the  extended plane of the Kerr geometries  are explored in Sec.\il(\ref{Sec:MBScharacteristics}).
Finally, we close this section with the analysis of some
limiting cases in Sec.\il(\ref{Sec:limit-acses}).
In Sec.\il(\ref{Sec:embo}), we  detail the \textbf{BH} thermodynamic properties in terms of  metric bundles.
 Sec.\il(\ref{Sec:disc-MBL})  discusses the role of  metric bundles in \textbf{BH} thermodynamics,  focusing on   the concept of  \textbf{BH} surface gravity, the first law of \textbf{BH} thermodynamics and  concluding with some notes on the \textbf{BH} transitions as described by the laws of \textbf{BH} thermodynamics.
Following this analysis,  in Sec.\il(\ref{Sec:nil-base-egi}), we explore \textbf{BH}  transitions in the extended plane.
The  \textbf{BH} irreducible   mass is the subject of Sec.\il(\ref{Sec:sud-mirr-egi}), where
 we examine the
    extraction of \textbf{BH}  rotational energy as constrained  with \MB s.
  Sec.\il(\ref{Sec:fin-BH-THEr-presenT}), \textbf{BH} thermodynamics in the extended plane, closes this section.
Finally, Sec.\il(\ref{Sec:final-remarks}) contains some concluding remarks.
%


\section{The Kerr-Newman metric}\label{Sec:axila}
Although in this chapter  we will focus mainly on the properties of the Kerr metric, it is useful to consider the Kerr geometry  as a limiting case of the more general Kerr-Newman solution.

The Kerr-Newman (\textbf{KN}) spacetime  is a solution of the Einstein--Maxwell equations, for an electro-vacuum, asymptotically flat spacetime   describing the geometry  surrounding a rotating, electrically charged mass, with  mass parameter $M$,  charge parameter $Q$, and dimensionless spin parameter $a\equiv J/M$ (the  {specific} angular momentum,  the rotational parameter associated to  the central singularity), while   $J$ is the
total angular momentum.

The line element in Boyer--Lindquist (BL) coordinates  $(t,r,\theta,\phi)$ is\footnote{For the seek of simplicity, where convenient, we adopt
geometrical  units with $c=1=G$.  The radius $r$ has unit of
mass $[M]$, and the angular momentum  units of $[M]^2$, the velocities  $[u^t]=[u^r]=1$
and $[u^{\phi}]=[u^{\theta}]=[M]^{-1}$ with $[u^{\phi}/u^{t}]=[M]^{-1}]$ and
$[u_{\phi}/u_{t}]=[M]$.  Latin and Greek indices run in $\{0,1,2,3\}$.}:
\bea\label{Eq:metric-KN}
ds^2=\frac{\sin^2\theta \left[d\phi \left(a^2+r^2\right)-a dt\right]^2}{\Sigma}-\frac{\Delta \left(a d\phi \sin^2\theta-dt\right)^2}{\Sigma}+\frac{ \Sigma dr^2}{\Delta}+\Sigma d\theta^2,
\eea
with    $r\in[0,+\infty)$, $t\in [0,+\infty)$, $\theta\in[0,\pi]$ and $\phi\in[0,2\pi]$, where $G=c=1$  and
\bea
\Delta\equiv a^2+Q^2+r^2-2M r\quad\mbox{and}\quad\Sigma\equiv a^2 \cos^2\theta+r^2.
\eea
In the following analysis, we shall use also the parameter $\sigma\equiv \sin^2\theta\in [0,1]$.
Let us introduce the total charge:
\bea
Q_T\equiv\sqrt{a^2+Q^2}.
\eea
 The \textbf{KN} metric describes black hole (\textbf{BH}) solutions for $Q_T\leq M$, with outer and inner horizons
 \bea
 r_{\pm}\equiv M\pm\sqrt{M^2-Q_T^2},
 \eea
 respectively
   in terms of the total charge.
   Extreme \textbf{KN}  \textbf{BHs}  are for  $Q_T=M$, where $r_+=r_-=M$.
   The metric (\ref{Eq:metric-KN}) describes
  \textbf{NSs} for $Q_T>M$.

The \textbf{KN} metric reduces to the Kerr solution for $Q=0$, where $Q_T=a$ (vacuum exact solution of the Einstein equations  describing  an axisymmetric,   stationary, asymptotically flat spacetime).

For $a=0$ the \textbf{KN} metric is the electrically charged,  spherically symmetric, asymptotically flat  Reissner--Nordstr\"om  (\textbf{RN}) solution, where $Q_T=Q$. The static, electrically neutral  and spherically symmetric  asymptotically flat  Schwarzschild  geometry  is the limit for $Q=0$ and $a=0$ (i.e. $Q_T=0$) where the \textbf{BH}   horizon is $r=2M$.

The ergosurfaces of the \textbf{KN} spacetimes
are defined by the zeros of the  norm of the  Killing vector field
$\xi_{(t)}^a=\partial_t$, which in BL coordinates are
\bea
r_\epsilon^\pm\equiv M \pm\sqrt{M^2-Q^2-a^2\cos^2\theta},
\eea
for the
outer and inner ergosurfaces, respectively (the Killing vector $\xi_{(t)}^a$ is space-like in the region $r\in]r_\epsilon^-,r_\epsilon^+[$).
In the spherically symmetric cases  ($a=0$),
we obtain  $r_\epsilon^\pm\equiv r_\pm$ and there are no ergosurfaces.

In the following sections, where convenient in complex expressions,  we   adopt  dimensionless quantities where $r\rightarrow r/M$ and $a\rightarrow a/M$.

\section{Light-surfaces  and metric bundles}\label{Sec:LS-MBS}
In this section, we build   the metric bundles  (\MB s)  framework using the properties of special light surfaces and the   \textbf{BH}  horizons.
We will introduce the concept of metric bundles on the basis of some spacetime light surfaces. We will illustrate some of their characteristics and relevance for    \textbf{BH} and \textbf{NS} spacetimes. In the next sections, we will precise   these definitions for the Kerr geometry, discussing  details and applications.

First introduced in  \cite{remnants}  for the Kerr geometries, framed   in the analysis of properties of the
 Kerr \textbf{BHs} and \textbf{NSs}, \MB s can be  generally defined in spacetimes with Killing horizons (for generalizations  to other spacetimes with Killing fields, see \cite{LQG,wormhole}).

The idea is to bundle geometries according to some particular characteristics common to all the geometries of the bundle.
This formalism, therefore,  allows  to explore   properties  in the  bundled metrics from a special,  ``global", perspective, studying spacetime properties as unfolding along the entire set of solutions. Bundles enlighten properties attributable to different points of spacetime and, on the other hand, connect different geometries, and in particular connect points of \textbf{BHs} and \textbf{NSs} spacetimes.
These properties can be  measured through the observation of special  light-like  orbits.

In this framework, we introduce the concept of
extended plane which, in brief,  can be defined as a  graphic representing the  collection of metrics  (\MB s curves)  related by a  common property.
  We specify below the details  of these definitions, while in the next sections we develop in details  the  set-up and applications.

\medskip

For the Kerr geometry, a metric Killing  bundle  is a collection of Kerr spacetimes characterized by  a  particular  photon circular  frequency, $\omega$ (bundles characteristic frequency), at which the norm of the stationary observers four-velocity  vanishes. We detail  this aspect in Sec.\il(\ref{Sec:Station-ob}).
It  is straightforward to prove that $\omega$ is also the frequency (angular velocity) of a  \textbf{BH} (inner or outer) horizon of the bundle.
A metric bundle  is
 represented by a curve in the  \emph{extended plane}, i.e.,  a plane $\mathcal{P}-r$, where $\mathcal{P}$ is a parameter of the  spacetime. In the Kerr geometry $\mathcal{P}$, for example, is the dimensionless spin, where $r$ is   the (dimensionless) radial Boyer-Lindquist  coordinate. Thus, an extended plane represents all the metrics of the same spacetime  family so that, by varying the parameter $\mathcal{P}$, we can extend a particular analysis to include all  metrics. In Figs\il(\ref{Fig:Plot-specif})--upper panel, we represent the extended plane of the Kerr spacetime on the equatorial plane $\sigma=1$.

Concentrating on the Kerr spacetime, the metrics of one metric Killing bundle with characteristic orbital  frequency $\bar{\omega}$ are all and the only Kerr  (\textbf{BHs} or \textbf{NSs}) spacetimes, where at some point $r$ the   light-like (circular) orbital frequency is $\omega=\bar{\omega}$. In  the extended plane, all the  curves associated to the \MB s (bundle curves) are tangent to the horizons curve (the  curve  representing in the extended plane   the  Killing horizons of all Kerr \textbf{BHs}). Then,  this tangency condition implies that  each bundle characteristic frequency $\omega$ coincides with the frequency   of a  \textbf{BH} Killing horizon. Consequently, the horizons in the extended plane emerge as the envelope surface of the collection of all the metric bundles curve--(Figs\il(\ref{Fig:Plot-specif})--upper panel).

The metric bundles of the Kerr spacetimes  contain either  \textbf{BHs} or \textbf{BHs} \emph{and}  \textbf{NSs}. Therefore, it is possible to find a  \textbf{BH-NS}  correspondence through the tangency condition of the \MB s to the horizons curve in the extended plane,  providing  also  an alternative  interpretation of \textbf{NSs} and  \textbf{BHs} horizons.

In fact, the exploration of \textbf{MBs}    singles out some fundamental properties of  the \textbf{BHs} and  \textbf{NSs} solutions, which are related, in particular, to the thermodynamic properties of \textbf{BH} spacetimes. We shall explore this aspect in Sec.\il(\ref{Sec:embo}).

The analysis of light surfaces defining  \MB s  opens  the possibility to extract information about the \textbf{BH} horizons, i.e., to  detect properties which are directly attributable  to  the presence  of a  \textbf{BH} horizon.
From Kerr metric bundles, we  define     the    ``horizons replicas"  special orbits  of a Kerr spacetime  with  photon (circular) orbital  frequency equal to the \textbf{BH} (inner \footnote{Note,  the inner horizon has been
shown to be unstable    and it has been conjectured it could
end  into a singularity
\cite{Simpenrose,DafermosLuk}.} or outer) horizon frequency, which coincides with  the  characteristic frequency of (one of the two) bundle(s) crossing the    point  on the extended plane $(r,a)$ corresponding to the replica orbit $r$ on of the spacetime with spin parameter $a$ (at fixed poloidal angle $\theta$)--Figs\il(\ref{Fig:PlotBri02m},\ref{Fig:PlotBri02mnsz},\ref{Fig:JirkPlottGerm1}).

  Therefore, replicas    could  be detected,  for example, from the spectra of electromagnetic emissions coming from \textbf{BHs} and, in particular, from locations close to   the \textbf{BH}  rotational axis--\cite{nuclear,GRG-letter}.

On the other hand, for  the
Kerr geometry it was pointed out  that  \textbf{NSs}  with  spin-mass ratio $a=J/M$  close to the value of the extreme \textbf{BH}, were  related, in the extended plane, to \MB s curves  tangent  to a portion of the inner horizon, and faster spinning \textbf{NSs}, with $a>2M$,  to the outer  horizons curve.
This property of the \textbf{NS} spacetimes in the extended plane explains a  peculiarity emerging in  some light surfaces.

	The light surfaces defining  \textbf{MBs},
 functions of  the   spin parameter $a$ and the  frequency  $\omega$,  bound a region in the plane $r-\omega$ for  \textbf{NS} spacetimes called Killing throats (tunnels).
	In \textbf{NS} geometries,  a  Killing  throat or tunnel
is a connected and bounded region in the $r-\omega$ plane,
containing  all the stationary observers allowed within
	two limiting   frequencies  $\omega\in]\omega_-, \omega_+[$. On the other hand, in the case of
	\textbf{BHs}, a Killing  throat is either a disconnected  region
	or a region  bounded by  singular  surfaces in the  extreme Kerr \textbf{BH} spacetime.
We detail this aspect in Sec.\il(\ref{Sec:light-surfaces-BottleN}).

Slowly spinning \textbf{NSs}  are characterized by  ``restrictions'', called  ``Killing bottlenecks", of the Killing tunnel,  which are related to metric bundles in the extended plane, through  the \MB{} curve tangency condition to the horizons curve \cite{remnants}.
The \textbf{BH} extreme Kerr spacetime, therefore,  represents the  limiting case of the  Killing bottleneck   (as defined  in the BL coordinates), where the tunnel narrowing  closes on the \textbf{BH} horizon--Figs\il(\ref{Fig:Plot-specif})--{lower} panel.
Killing bottlenecks were   also connected with the concept of  pre-horizon regime introduced in
\cite{de-Felice1-frirdtforstati,de-Felice-first-Kerr}.
 The pre-horizon regime was analyzed in \cite{de-Felice-first-Kerr}.
It was shown
that a gyroscope would conserve a  ``memory" of the static or
stationary initial state  \cite{de-Felice3,de-Felice-mass,de-Felice4-overspinning,Chakraborty:2016mhx}.
Adopting a similar idea of plasticity, bottlenecks have also been named  horizon remnants.
Killing throats and bottlenecks    were also grouped in \cite{Tanatarov:2016mcs}  in   structures named ``whale  diagrams'' of the Kerr and Kerr-Newman spacetimes--see also
\cite{Mukherjee:2018cbu,Zaslavskii:2018kix}.

Below,  we precise the  \textbf{MBs} definitions  for Kerr spacetimes and relate them explicitly to quantities of importance in \textbf{BH}  thermodynamics.
We   discuss the relations between \textbf{MBs}, stationary observes, and light surfaces, which are used to constrain  many processes associated to the physics of jet emission, accretion disks, and  energy extraction from \textbf{BHs}.

\medskip

In details,
in Sec.\il(\ref{Sec:majo-sure}), we discuss the Killing horizon definitions in relation to metric bundles.
Stationary observers and metric bundles are the focus in Sec.\il(\ref{Sec:Station-ob}).
Light-surfaces and bottlenecks are investigated in Sec.\il(\ref{Sec:light-surfaces-BottleN}).
\MB s characteristics in the Kerr extended plane are explored in Sec.\il(\ref{Sec:MBScharacteristics}).
Finally, we close this section with the analysis of some
limiting cases in Sec.\il(\ref{Sec:limit-acses}).

\subsection{Killing horizons}\label{Sec:majo-sure}
	Let us introduce the   Killing vector $\mathcal{L}^a=\xi^a_{(t)} +\omega \xi_{(\phi)}^a$ with
 $\xi_{(t)}^a=\partial_{t} $ representing the stationarity of the  background  and the  rotational  Killing field
 $\xi^a_{(\phi)}=\partial_{\phi} $ of the KN spacetime. The quantity  {$\mathbf{\mathcal{L_N}}\equiv g(\mathcal{L},\mathbf{\mathcal{L}})$} becomes null  for photon-like
particles with orbital  frequencies $\omega=\omega_{\pm}$.

 The  limiting   frequencies (or relativistic velocities) $ \omega_H^\pm\equiv\lim_{r\to r_\pm}\omega_{\pm}$ are the outer and inner  \textbf{BH}  horizons frequencies (relativistic angular velocities),  representing  the \textbf{BH} rigid rotation.

 Thus, the null frequencies  $\omega=\omega_\pm$  evaluated on the \textbf{BH} horizons, are the \textbf{BH} inner and outer horizons frequencies, $\omega_H^\mp$, respectively.
 There is $\omega_H^-\geq \omega_H^+$, where
 $\omega_H^-=\omega_H^+=1/2$ for the  extreme Kerr \textbf{BH}.

The vector fields  $\mathcal{L}_H^{\pm}\equiv \mathcal{L}(r_{\pm})=\xi_{(t)} +\omega_H^{\pm} \xi_{(\phi)}$ define,  therefore, the
\textbf{BH} horizons   as   Killing horizons\footnote{The  Killing vector
$
\mathcal{L}^a_{\pm}\equiv \xi_{(t)}^a+\omega_{\pm}\xi_{(\phi)}^a
$,
where $\omega_{\pm}$ satisfies the null condition on the norm of $\mathcal{L}^a$,
can be interpreted as generator of  null curves ($g_{ab}\mathcal{L}^a_{\pm}\mathcal{L}^b_{\pm}=0$)
as the Killing vectors $\mathcal{L}_H^{\pm} $ are also generators of Killing   horizons.}.

The   Killing horizons are  {null} (light--like) hypersurfaces generated by the flow of a Killing vector,
whose null generators coincide with the orbits of a
one-parameter group of isometries, i.e., in general,    there exists a Killing field $\mathcal{L}$, which is normal to the null surface.
The Kerr  \textbf{BH} horizon $r_+$ is,  therefore,  a non-degenerate (bifurcate) Killing
horizon generated by the vector field $\mathcal{L}$.
In the case $a = 0$ (the Schwarzschild and \textbf{RN}  spacetimes, where $\omega=0$), the Killing vector
$\xi_{(t)}$, generator of $r_+$, is hypersurface-orthogonal\footnote{The (strong)
rigidity theorem connects the event horizon with a Killing
horizon--see \cite{HE,FRW}.
The event horizons  of a spinning \textbf{BH}  are   Killing horizons   with respect to  the Killing field
$\mathcal{L}^\pm_H=\xi_{(t)}+\omega_H^{\pm} \xi_{(\phi)}$, where  $\omega_H^{\pm}$ is the angular velocity (frequency) of the horizons,  representing   the \textbf{BH} rigid rotation. Or, the event horizon of a stationary asymptotically flat solution with matter satisfying suitable hyperbolic equations  is a Killing horizon.
In the limiting case of spherically symmetric, static  spacetimes,
 the event horizons are  Killing horizons with respect to the  Killing vector
$\partial_{t}$ and  the
event, apparent, and Killing horizons  with respect to the  Killing field   $\xi_{(t)}$ coincide (we can say that $\omega_H^+=0)$.
}.

Metric Killing bundles or metric bundles   are structures defined as solutions of $\mathcal{L}_{\mathcal{N}}\equiv g(\mathcal{L},\mathcal{L})=0$ with \emph{constant} $\omega$. Therefore, they
 are  the collections of all and only geometries defined as
  solutions of  $g(\mathcal{L},\mathcal{L})=0$,   having photon circular orbits with equal  (constant)  orbital frequency $\omega$, called \emph{bundle characteristic frequency}.
   It  has been proved that (Kerr) metric Killing bundles always contain at least one \textbf{BH} geometry.
     A bundle can contain \textbf{BHs} or \textbf{BHs} \emph{and} \textbf{NSs} solutions.

Notably, many quantities considered in this analysis are conformal invariants of the metric
and inherit some of the properties of the Killing vector $\mathcal{L}$ up to a conformal transformation.
The simplest case is when one considers a conformal expanded (or contracted) spacetime  where $\tilde{\mathcal{L}}_{\mathcal{N}}\equiv
\tilde{g}(\mathcal{L}, \mathcal{L})=\Omega^2 g(\mathcal{L}, \mathcal{L})=\Omega^2 \mathcal{L}_{\mathcal{N}}
$.
This holds also   for a   Killing tensor $\tilde{\mathcal{L}}\equiv \Omega \mathcal{L}$.

  Focusing our discussion on the \textbf{KN} and Kerr  spacetimes, the bundle characteristic frequency coincides    (in magnitude) to an (inner or outer)  \textbf{BH}  horizon frequency, $\omega_H^{+}$ or $\omega_H^-$, of the bundle.

Kerr \MB s  can be found, for example,   as  solutions  for the dimensionless spin  $a:\mathcal{L}_\mathcal{N}=0$ for each fixed values of the constant $\omega$.
We can   consider the co-rotating,  $a\omega>0$, and  the counter-rotating, $a\omega<0$,  orbits  separately with   frequencies that
are equal in magnitude  to the horizon frequencies. Nevertheless, here we will consider mainly  $a\geq 0$ and $\omega\geq0$.

Below, we explore in details the definitions  of    \MB s in relation to stationary observers and horizons replicas, introducing the concept of extended  plane allowing the \MB s representation as curves tangent to the \textbf{BH} horizon curves in the plane. We also investigate metric bundles in the Schwarzschild spacetime  as limits of  bundles of  the Kerr and \textbf{RN} spacetimes.

\subsection{Stationary observers and metric bundles}\label{Sec:Station-ob}
{The definition of \textbf{MBs}} is tightly  connected  to the definition of stationary observers, whose  worldlines have
 a tangent vector which is  a  Killing vector. Hence, in the Kerr spacetime, for example,
 their    four-velocity  $u^a$ is    a
linear combination of the two Killing vectors $\xi^a_{(\phi)}$ and $\xi_{(t)}^a$, i.e.{
$
u^a=\Omega\mathcal{L}^{a}= \Omega (\xi_{(t)}^a+\omega \xi_{(\phi)}^a$)},  where
 $\Omega$ is a normalization factor and $d\phi/{dt}={u^{\phi}}/{u^t}\equiv\omega$ is the stationary observer relativistic angular velocity\footnote{Because of the spacetime  symmetries, the coordinates  $r$ and $\theta$  of a stationary observer are constants along  its worldline, i. e.,  a stationary observer does not see the spacetime changing along its trajectory. Static observers are defined by the limiting condition  $\omega=0$  and  cannot exist in the ergoregion.}.
The causal structure defined by timelike stationary  observers is characterized by a frequency   bounded in the range $\omega\in]\omega_-,\omega_+[$ (\cite{malament}).
The limiting frequencies  $\omega_{\pm}$ are then  solutions of the condition $\mathcal{L_{N}}=0$, determining
the frequencies $\omega_H^{\pm}$ of the Killing horizons, which coincide with the metric bundles characteristic frequency.

\MB s provide, therefore,  the limiting orbital frequencies of stationary observers,  constraining  the
 timelike (matter circular) motion.

\subsection{Light surfaces and bottlenecks}\label{Sec:light-surfaces-BottleN}

\textbf{Light surfaces}

For Kerr spacetimes,  the solutions
$a: \mathcal{L}_{\mathcal{N}}=0$ for fixed constant $\omega$ define the \MB s with characteristic frequency $\omega=$constant. For a fixed Kerr spacetime ($a/M=$constant), the solutions
$\omega=\omega_\pm: \mathcal{L}_{\mathcal{N}}=0$ define the spacetime  stationary observers limiting frequencies  $\omega_\pm$ as functions of  the orbit $r$.
Also, for a fixed Kerr spacetime ($a/M=$constant), the  solutions
$r=r^\pm_s: \mathcal{L}_{\mathcal{N}}=0$, define the light surfaces of the Kerr spacetimes having photon-like (circular orbiting) frequencies $\omega_\pm$.
Physical (timelike) stationary  observers are located between the two light surfaces $]r_s^-,r_s^+[$ with orbital frequencies in the range $\omega\in]\omega_-,\omega_+[$ (where  $\omega_-<\omega_+$).
We have that  $\omega_+=\omega_-$ on $r=0$ and
$r=r_\pm$.
Therefore, at each point of the extended plane,
there is a maximum of two \MB s curves crossing with characteristic frequency $\omega_+$ and $\omega_-$, respectively.

In other words, \MB s  are related to the light surfaces  $r^\pm_s$  in a fixed spacetime,  defining the photon orbits $r=r^\pm_s$  with  orbital frequencies $\omega_\pm$, constraining the stationary observers. \MB s have characteristic frequencies  $\omega_\pm$,  therefore,
\MB s relate different geometries through their light surfaces $r_s^\pm$.

\textbf{Killing tunnels and Killing bottlenecks}

Slowly spinning Kerr  \textbf{NSs} are characterized by the emergence  of special structures  in their light surfaces,  the  Killing bottlenecks, which can be seen through  \MB s.

\begin{figure*}
\centering
       \includegraphics[width=9cm]{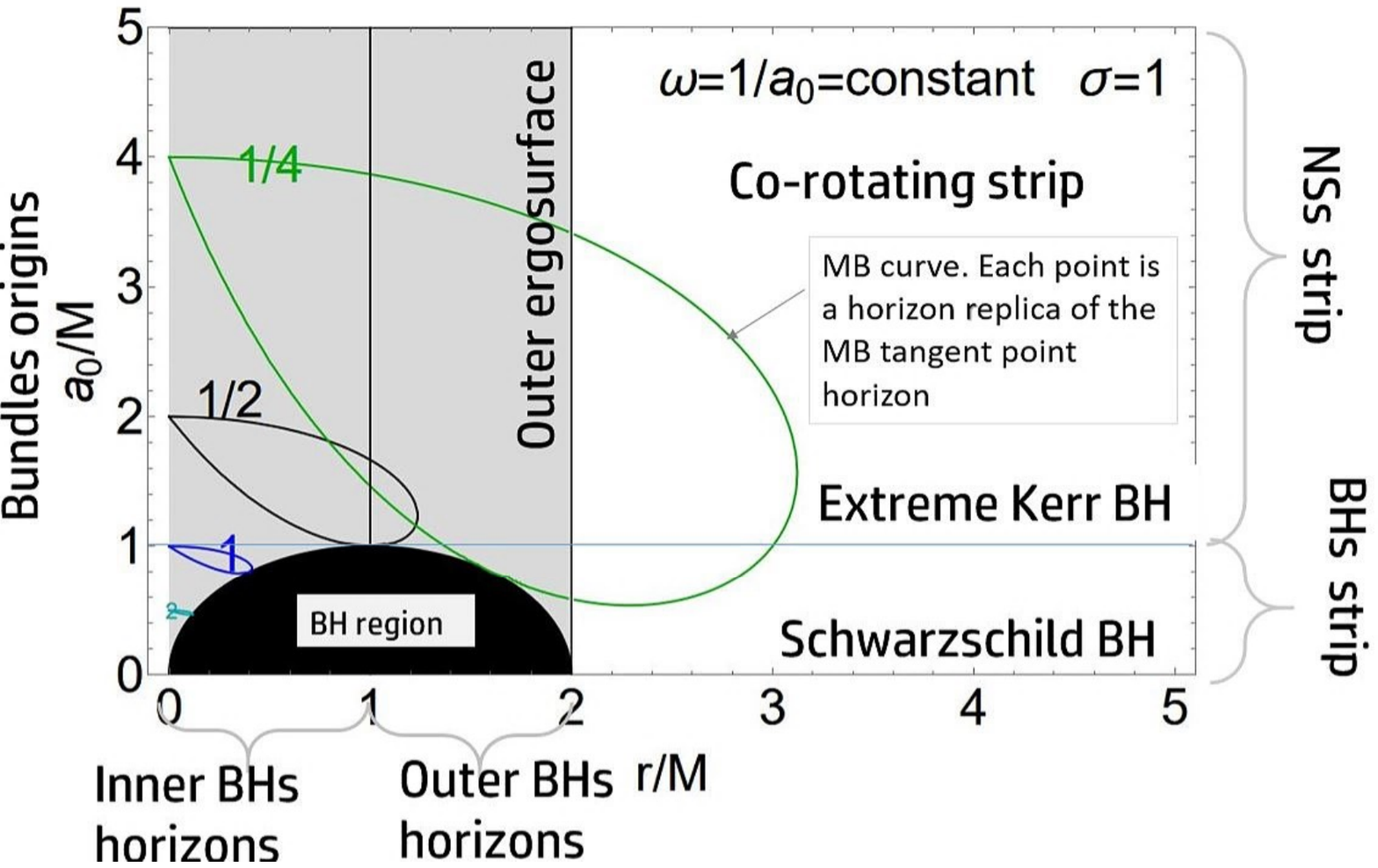}
        \includegraphics[width=9cm]{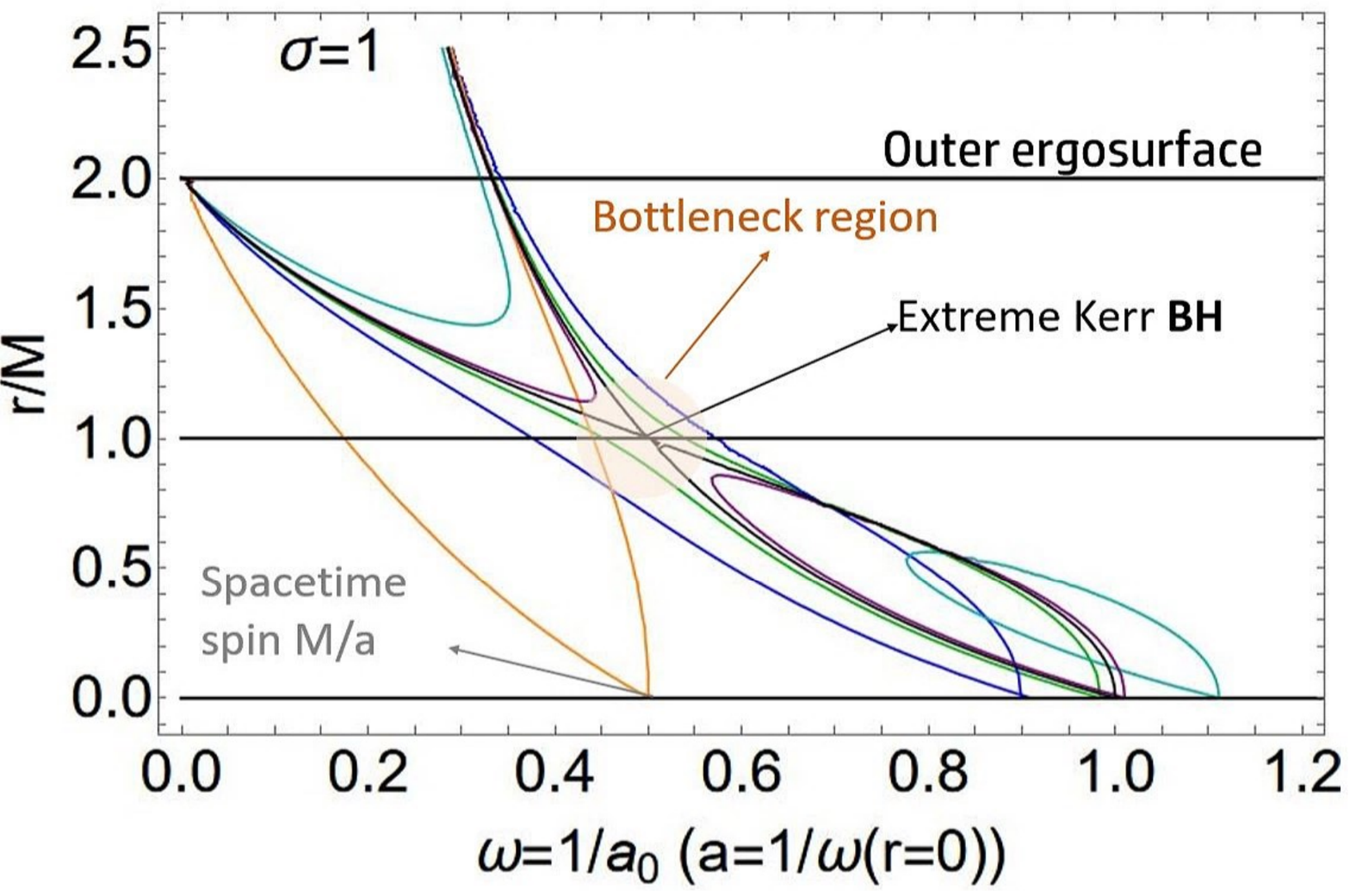}
           \caption{Upper panel: Extended plane $a/M-r/M$  on the equatorial plane $(\sigma=\sin^2\theta=1)$ of the Kerr geometry. Metric bundles (\MB s)  with   characteristic frequency $\omega=1/a_0=$constant  are shown  for different values of $\omega$, where $a_0$ is the bundle origin spin (bundles value at $r=0$). The black region  represents the \textbf{BH} region with  the outer and inner horizons $a_\pm$, as functions of  $r/M$.
A horizontal line on the extended plane at $\sigma=1$ represents a fixed spacetime  $a/M=$constant. In particular,
$a=0$ corresponds to the Schwarzschild \textbf{BH} spacetime and $a=M$ to the extreme Kerr \textbf{BH}. Naked singularities correspond to values $a>M$. The extended plane is shown for co-rotating orbits, i.e., for $a_0\geq0$ and $\omega\geq 0$. The frequencies of the inner and outer horizons curve are clearly distinguished.  The frequency $\omega^{\pm}_H=1/2$ corresponds  to the extreme Kerr \textbf{BH}. On the equatorial plane, the point $r=2M$ is the outer ergosurface (the  Schwarzschild \textbf{BH} horizon for $a=0$).
{Lower} panel:  Light surfaces $r_s^{\pm}(\omega;a)$ on the equatorial plane, solutions of  condition
$\mathcal{L}_{\mathcal{N}}=0$, as functions of the frequency $\omega$, for different spacetimes. Here, $\omega=1/a_0$ corresponds to $r=0$.  The extreme \textbf{BH} spacetime and the bottleneck region are also pointed out.}\label{Fig:Plot-specif}
\end{figure*}

Let us concentrate on   the plane $r-\omega$ for $\sigma=1$ (equatorial plane), of the \textbf{NSs}  geometries as represented in Figs\il(\ref{Fig:Plot-specif})--{lower} panel. At fixed spin $a>M$, for each Kerr spacetime,   the regions  bounded by the light surfaces $r_s^\pm$ are  called Killing throats or tunnels.
Stationary  observers orbits  of  the Kerr \textbf{NSs} are defined in the  Killing throats.

In the Kerr \textbf{BH}  spacetimes, there is $r_s^-=r_s^+$ on the  singularity $r=0$ and on the horizons  $r_\pm$.
The surfaces, $r_s^\pm$ in Figs\il(\ref{Fig:Plot-specif})--{lower} panel, merge  for  $a=M$, the extreme Kerr spacetime, where $r_s^\pm=r_\pm=M$.
Increasing the spin $a>M$, the Killing throat is defined by the      connected surfaces $r_s^\pm$.

The Killing bottleneck
is  as restriction of the  Killing tunnel (and in particular of the orbital range bounded by  $r_s^\pm$) boundaries for \textbf{NSs}, for spins  $a=]0,2M]$, close to the orbits $r=M$.
Killing
bottlenecks, therefore,  are restrictions
of the Killing throats appearing in special classes of slowly
spinning \textbf{NSs} and they are related to the \MB s features   and the  ergoregions, where  repulsive gravity effects appear.
In this region, the stationary observer orbital range  restrict to a (non zero) minimum range,
which is null in the limit of  extreme \textbf{BHs}.

Similar  structures  have been recognized  in
\cite{de-Felice1-frirdtforstati,de-FeliceKerr,de-Felice-anceKerr,de-Felice-first-Kerr}, by using the concept of pre-horizons, introduced in \cite{de-Felice1-frirdtforstati}.
 There is a pre-horizon regime in the
spacetime when there are mechanical effects allowing circular orbit observers which can recognize the close presence
of an event horizon \cite{de-Felice3,de-Felice-mass,de-Felice4-overspinning}.

The concept of remnants, bundles  with close characteristic frequencies values,  evokes  a sort of spacetime ``plasticity'' (or memory),
which we will deepened in the next section introducing the concept of extended plane.
(See also \cite{Chakraborty:2016mhx}, for a discussion on similar surfaces).

 \begin{figure*}
\centering
\includegraphics[width=5.75cm]{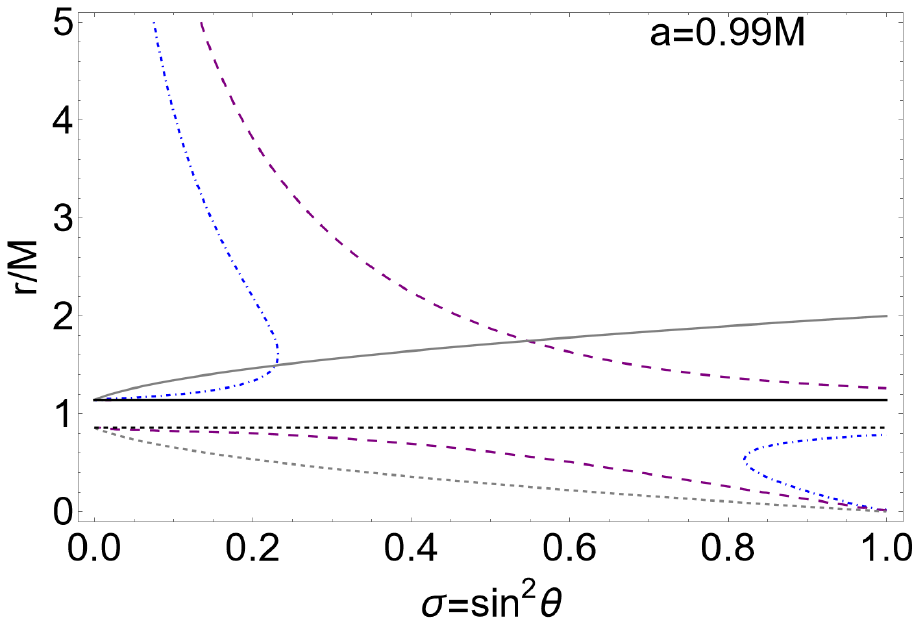}
      \includegraphics[width=5.75cm]{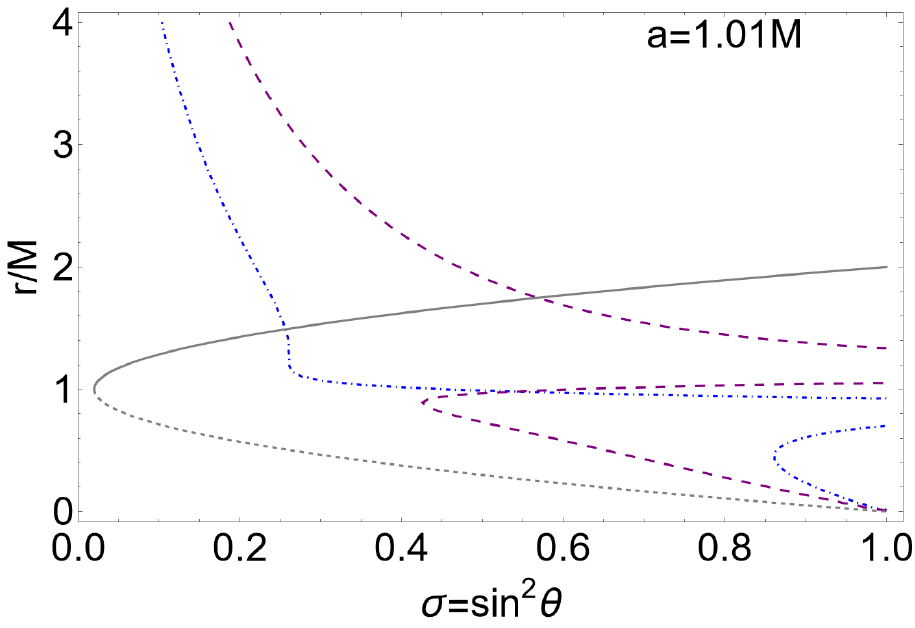}
     \includegraphics[width=5.75cm]{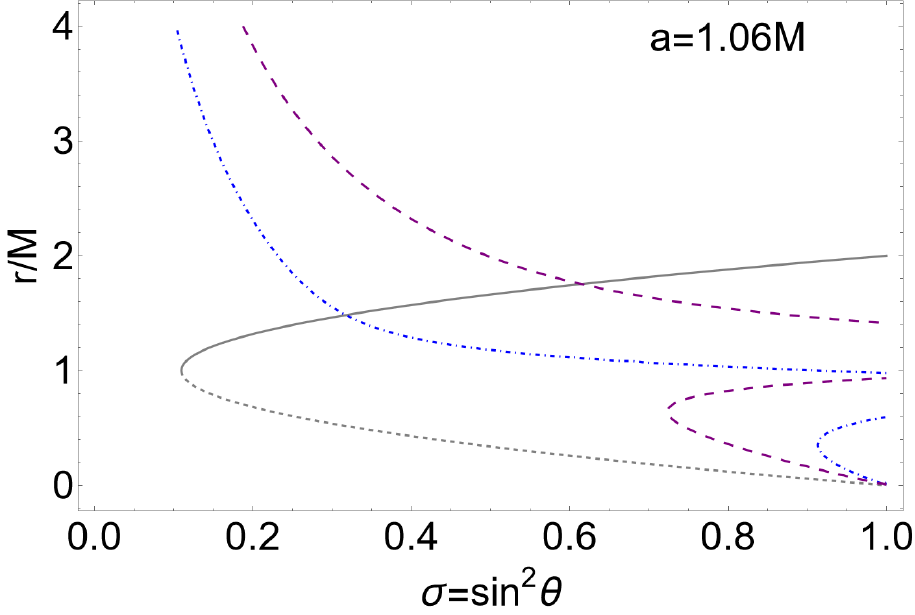}
      \includegraphics[width=5.75cm]{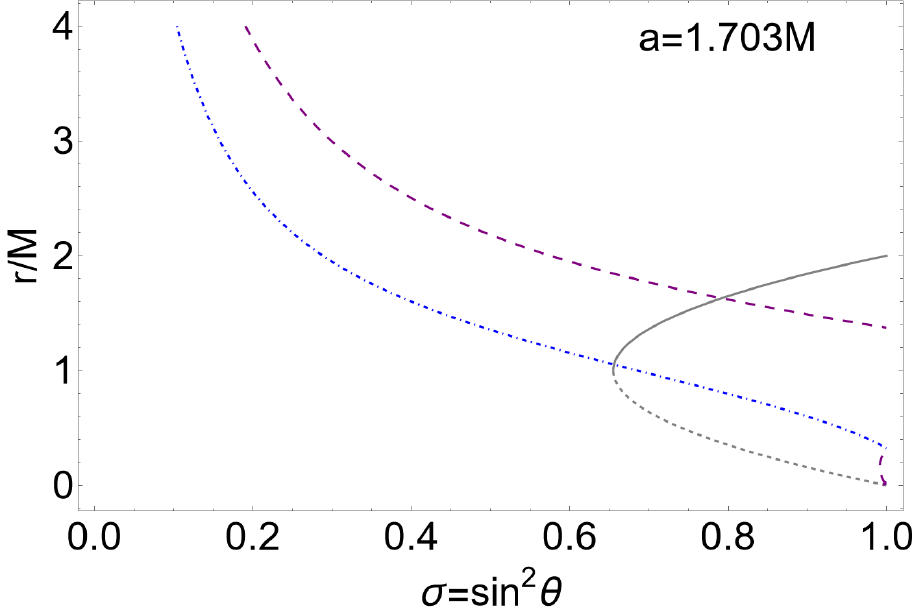}
 \includegraphics[width=5.75cm]{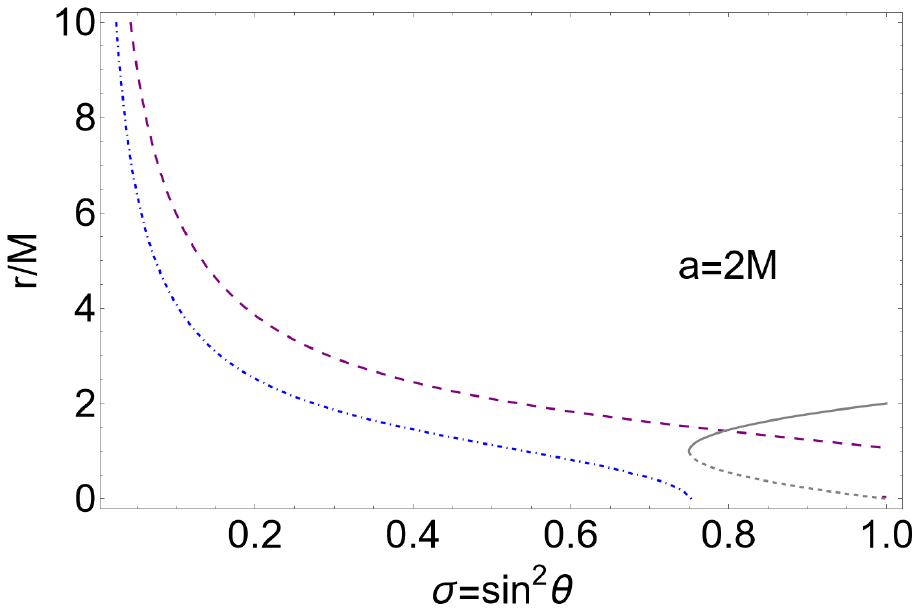}
        \includegraphics[width=5.75cm]{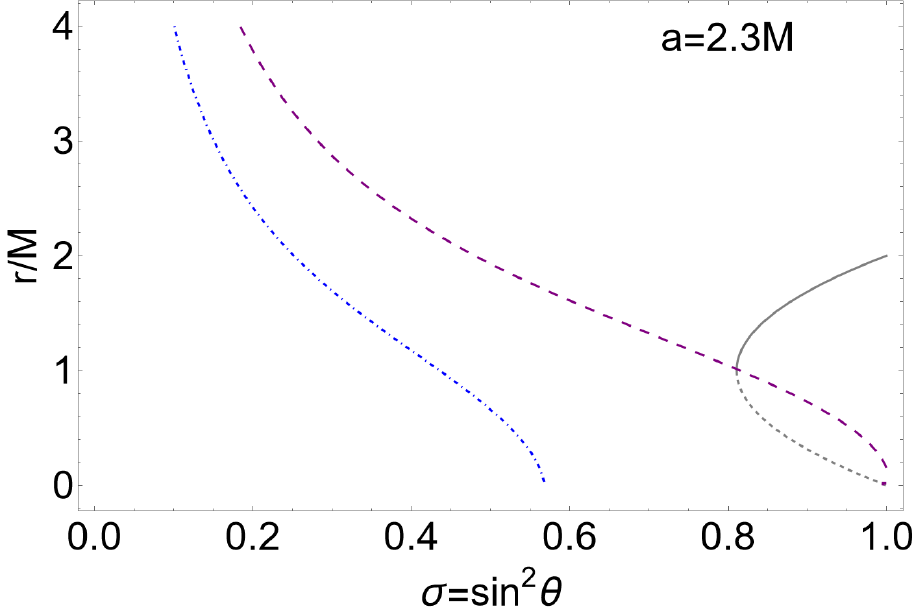}
        \caption{Co-rotating replicas  in  the  Kerr geometry with  dimensionless spin $a/M$.   Panels show  the replicas radius $r$ on each plane $\sigma\equiv\sin^2\theta\in [0,1]$ for the \textbf{BH}  with spin $a=0.99M$ in the \textbf{BH}  spacetime (upper left panel) and in different \textbf{NSs} spacetimes with spin $a/M$ signed on the panels. {Lower} right panel is a close-up view of the {lower} left panel.  Gray solid (dashed) curve is the outer  (inner) ergosphere $r_\epsilon^+(\sigma)$ ($r_\epsilon^-(\sigma)$).  {Purple large-dashed} ({blue dotted-dashed}) curves are the outer (inner) horizons  co-rotating replicas for the \textbf{BH}  with spin $a=0.99M$. See also Figs\il(\ref{Fig:PlotBri02mnsz}) and Figs\il(\ref{Fig:JirkPlottGerm1}).}\label{Fig:PlotBri02m}
\end{figure*}
 \begin{figure*}
 \includegraphics[width=6cm]{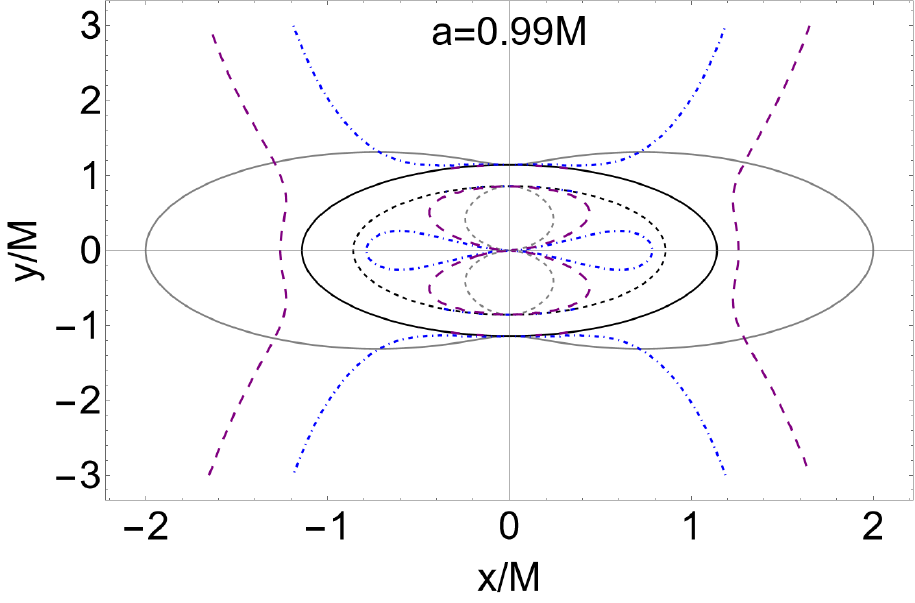}
 \includegraphics[width=6cm]{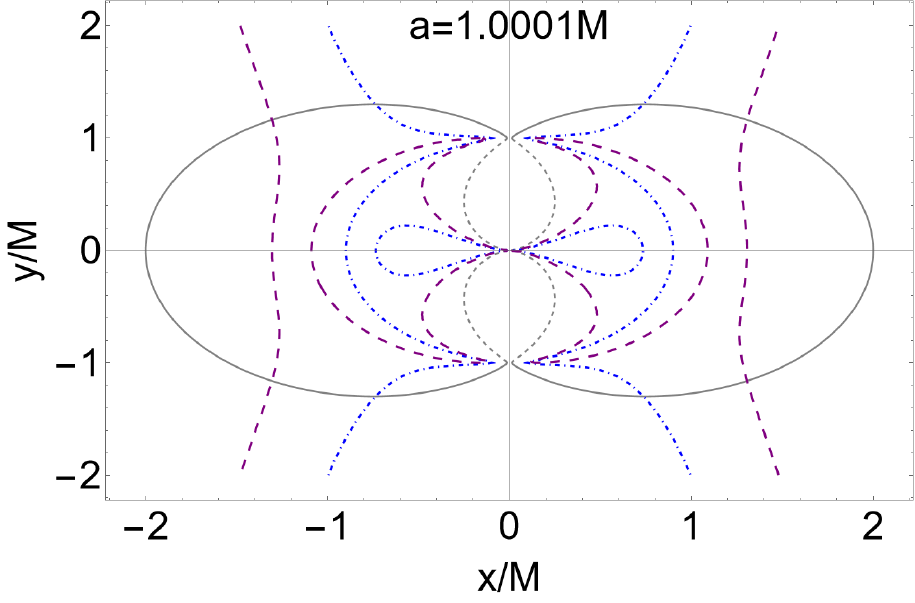}
   \includegraphics[width=6cm]{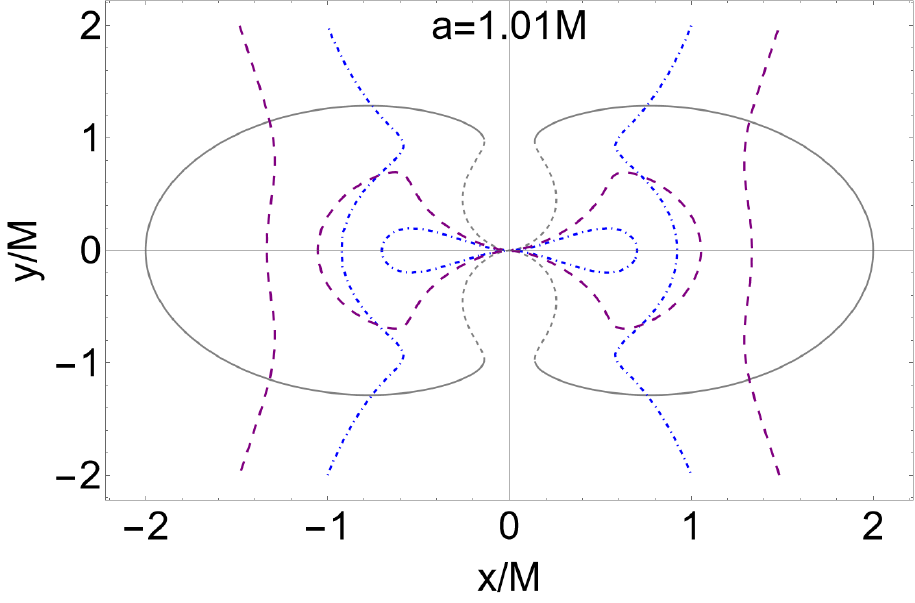}
     \includegraphics[width=6cm]{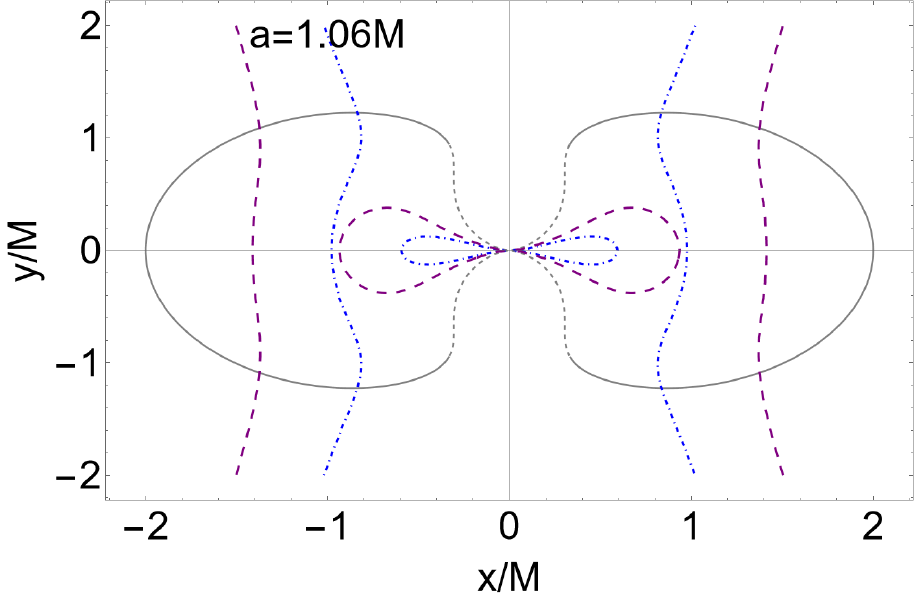}
       \includegraphics[width=6cm]{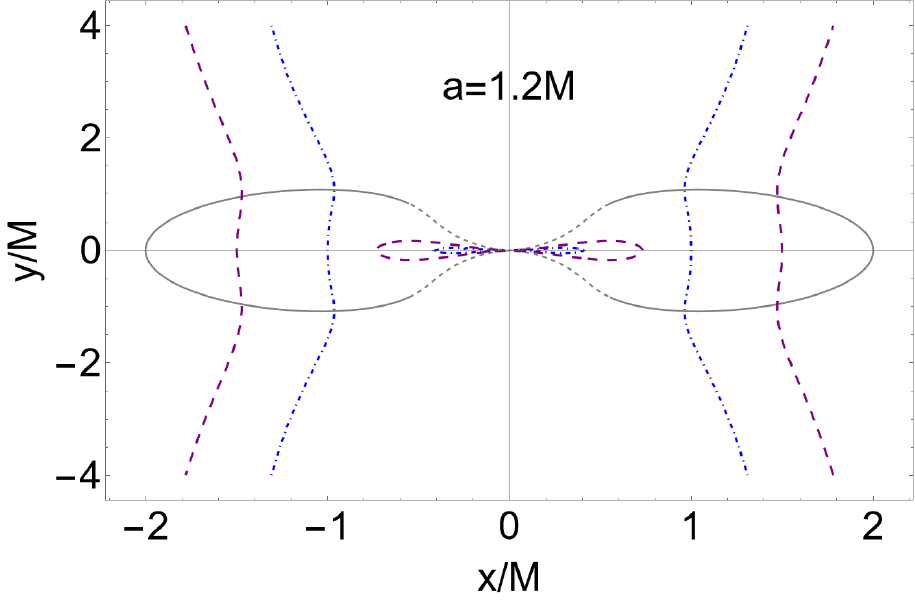}
      \includegraphics[width=6cm]{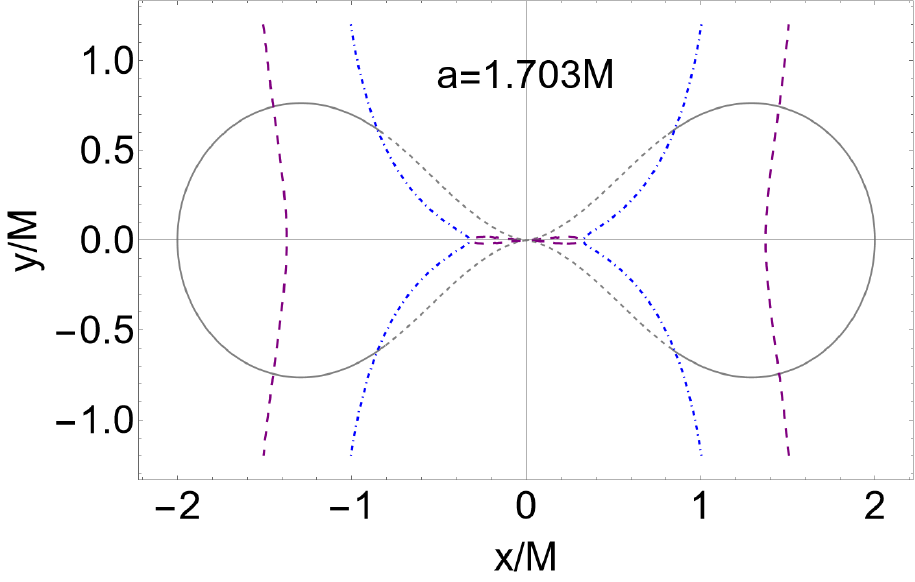}
      \includegraphics[width=6cm]{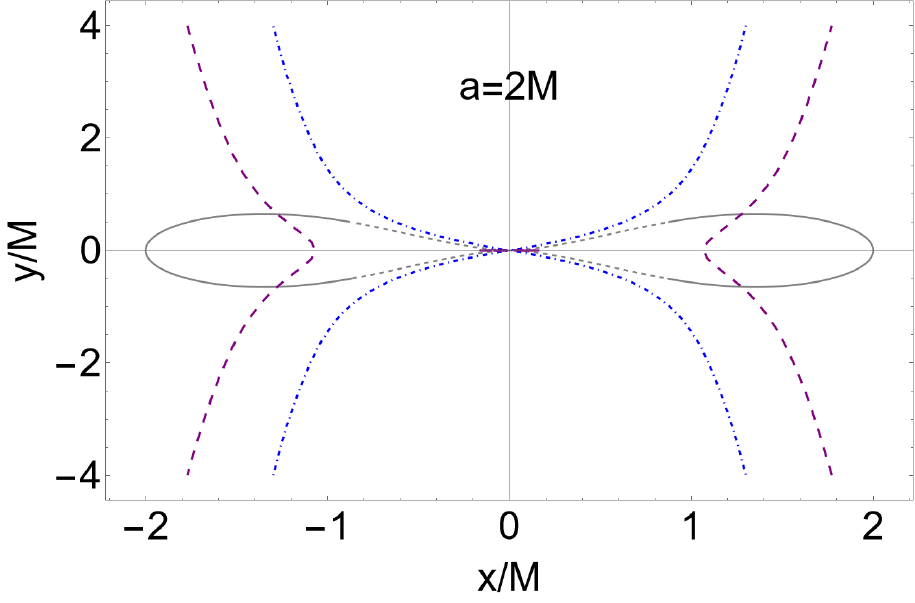}
          \includegraphics[width=6cm]{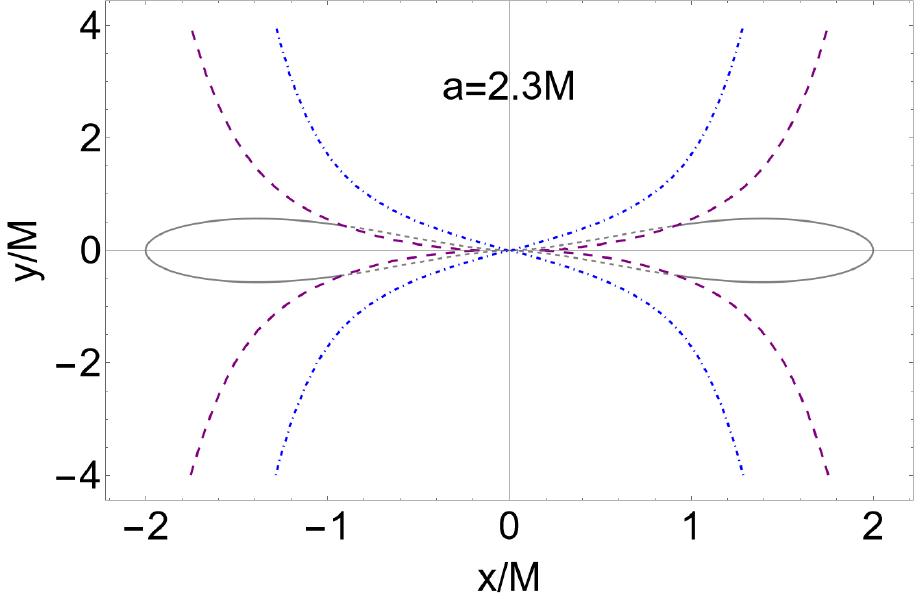}
 \caption{Co-rotating replicas of the \textbf{BH} horizons represented in  Fig.\il(\ref{Fig:PlotBri02m}) for different values of $a/M$ corresponding to \textbf{BH} s  with spin $a=0.99M$ and \textbf{NSs}  with different spin--mass rations $a/M$ as signed on the panels.  Gray solid (dashed) curve is the outer (inner) ergosurface. Black solid (dashed) is the outer (inner) \textbf{BH}  horizon. {Purple large-dashed (blue dotted-dashed)} curve  is the outer (inner) \textbf{BH}   horizon  co-rotating replica.
 (Here $x= r\sin\theta$ and $y=r\cos\theta$). See also Figs\il(\ref{Fig:PlotBri02m}) and Figs\il(\ref{Fig:JirkPlottGerm1}).}\label{Fig:PlotBri02mnsz}
\end{figure*}
 \begin{figure*}
\centering
 \includegraphics[width=9cm]{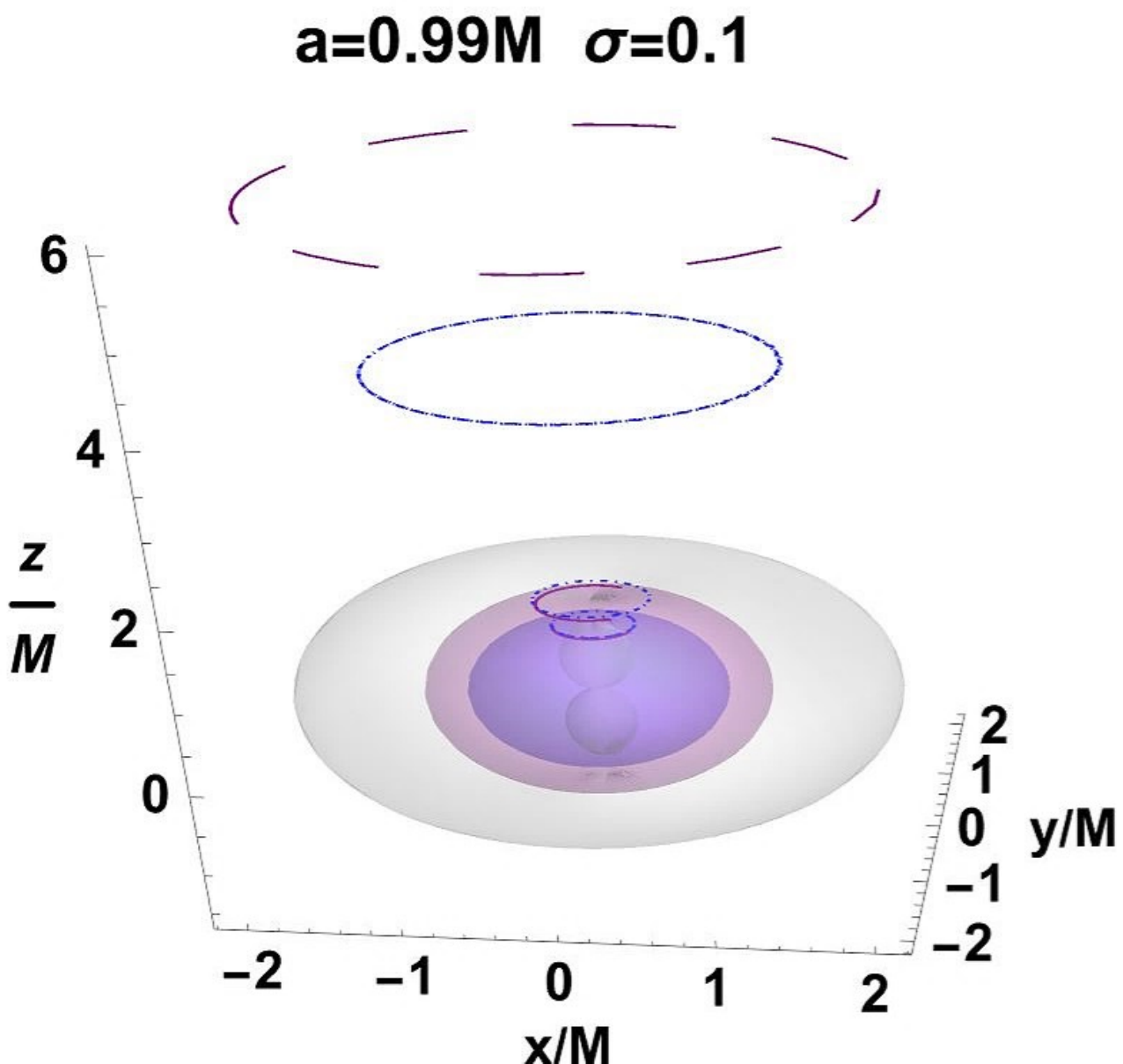}
  \includegraphics[width=9cm]{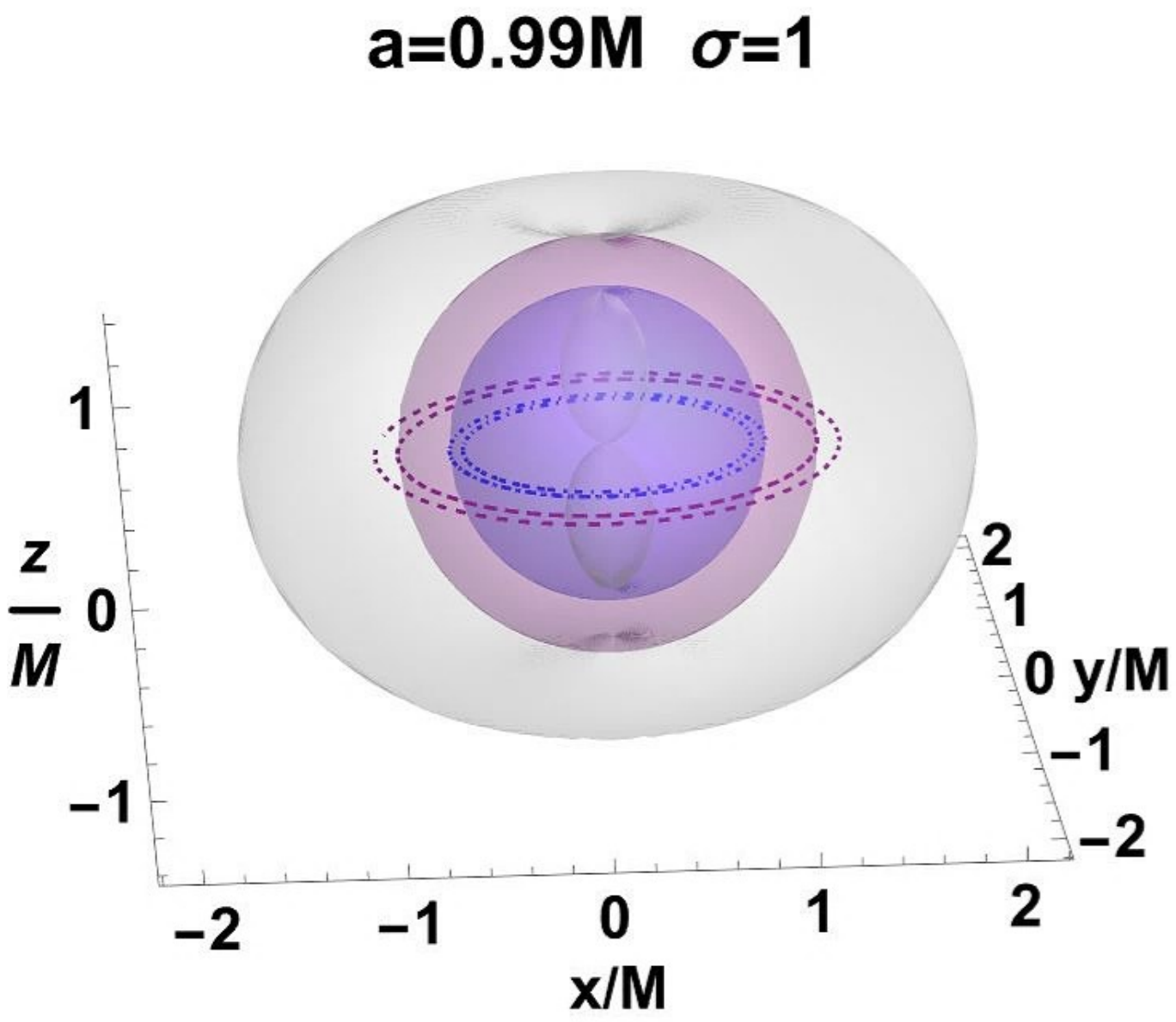}
 \caption{Co-rotating replicas of the \textbf{BH} horizons represented in  Fig.\il(\ref{Fig:PlotBri02m}) for the \textbf{BH}  with spin  $a/M=0.99$  on the plane $\sigma=0.1$ (with $\sigma\equiv\sin^2\theta\in[0,1]$)--upper panel-- and on the equatorial plane $\sigma=1$--{lower} panel.  {Purple large-dashed (blue dotted-dashed)}  curve is the outer (inner) \textbf{BH}  horizon co-rotating replicas. Here,
 $\{z=r \cos\theta, y=r \sin\theta \sin\phi, x=r \sin\theta\cos\phi\}$. The central singularity is
   the point $(x=0,y=0,z=0)$.   Gray surfaces are the  outer and inner ergosurfaces. Blue  surface is the inner horizon, purple surface is the outer horizon. See also Figs\il(\ref{Fig:PlotBri02m}) and Figs\il(\ref{Fig:JirkPlottGerm1}).}\label{Fig:JirkPlottGerm1}
\end{figure*}
\subsection{\MB s characteristics in the Kerr extended plane}\label{Sec:MBScharacteristics}

Below we list some of the main \MB s characteristics in the extended plane.
The concept of metric bundles   and some of their  main features are illustrated  in Fig.\il(\ref{Fig:Plot-specif})--upper panel, using  the extended plane  $a/M-r/M$ of the  Kerr  geometry on the equatorial plane $(\sigma=\sin^2\theta=1)$.

\medskip

Metric bundles of the Kerr spacetimes are solutions of  $\mathbf{\mathcal{L_N}}=0$, expressed as  functions of $(\omega,r,\sigma)$ and can be represented  as curves of a plane $a-r$ or $\la-r$ (\emph{extended plane}), where $\la\equiv a\sqrt{\sigma}$ is the origin spin for any plane $\sigma$--\cite{remnants,bundle-EPJC-complete,LQG}. Figs\il(\ref{Fig:Plot-specif})--upper panel represents the extended plane  of the Kerr geometry for $\sigma=1$, where  the Kerr bundle curves can be represented through the  (dimensionless) spin-function:
\bea\label{Eq.lcospis}
a(r,\omega)=\frac{2 \omega \pm\sqrt{r^2  \omega ^2 \left[1-r (r+2)  \omega ^2\right]}}{(r+1)  \omega ^2}.
\eea

The black region is  the \textbf{BH} region  bounded  by  the outer  $(r=r_+\in[M,2M])$ and inner $(r=r_-\in]0,M])$ horizons curves  $a_\pm\in[0,M]$ of the extended plane $a-r$.
Bundles are not defined in the  region $]r_-,r_+[$  of \textbf{BH} spacetimes, but they are defined in the \textbf{BH}  region $r\in[0,r_-]$ and in $r\in [0,2M]$ of \textbf{NS}.
The point $(a=0, r=0)$ is the  singularity, $(a=M,r=M)$, the maximum of the curve $a_{\pm}$, is the horizon of the extreme Kerr \textbf{BH}, $(a=0, r=2M)$ is the horizon of the Schwarzschild \textbf{BH} . The radius $r=2M$
 in the extended plane  coincides  also with the outer ergosurface  on the equatorial plane of all the  Kerr \textbf{BHs} and \textbf{NSs}.
Each line $a_{\pm}=$constant in this extended plane  is a \textbf{BH} geometry.  In particular $a=0$ to the Schwarzchild geometry and  the zero \MB s.
Per definition the   bundle characteristic  frequency $\omega$
is constant along the  \MB s curve,  in particular,
the frequency of the origin $r=0$ coincides with the bundle frequency $\omega$.

At a point $ r$, in
general, there are two different limiting photon frequencies $\omega_\pm$ for the stationary observers;
then, it follows that at each point of the extended plane (at fixed $\sigma$ and  with the
exception of the horizon curve) there have to be a maximum of
two different crossing metric bundles.

Each \MB\phantom{}  curve is tangent to the horizon curves in the extended  plane at one point only, defining,  therefore,  uniquely the \textbf{BH} with the \MB\phantom{}  frequency $\omega$,
coinciding   with the frequency of the horizon
at the  point where the bundle is tangent to the horizon curve.
The fact that metric bundles are tangent to the horizons curve in the extended plane has  significant implications.

The tangent point  distinguishes  the tangent spin  $a=a_g\in[0,M]$, on the curve $a_\pm$, but it is defined by the tangent radius $r_g\in[0,M]$ on inner horizons curve or  $r_g\in [M,2M]$  on the outer horizon curve.
The horizon curve emerge as the envelope surface of all \MB s.

Bundles can contain either only \textbf{BHs} or  \textbf{BHs} and \textbf{NSs}.
In the extended plane, \textbf{NSs} are ``necessary'' for the  construction of horizons curve. The outer  horizons curve emerges as envelope surface of bundles  with origin in \textbf{NSs}.
Part of the inner horizons curve (on the equatorial plane) emerges from \MB s with origin in the slowly spinning \textbf{NSs}. Bottlenecks in these \textbf{NSs} spacetimes coincide in the extended plane with the region  where these
\MB s curves are defined.
Killing bottlenecks appear in the \textbf{NSs} region of the extended plane containing parts of the \MB s tangent to the  inner horizons.
(The light surface  $r_s(\omega):\mathcal{L}_{\mathcal{N}}=0$
emerge as the  collections of all points, crossing of the \MB{} curve with a line $a=$constant on the extended plane.)

\textbf{MB}  curves in the extended plane connect points of different  (\textbf{BH} or \textbf{BH} and \textbf{NS}) geometries  having all the same characteristic null frequency $\omega$, which is the replica of a \textbf{BH}  horizon.
Fig.\il(\ref{Fig:Plot-specif})  shows  metric bundles at $\sigma=1$   for different   characteristic frequencies $\omega=1/a_0=$constant, where $a_0$ is the bundle origin on the plane $\sigma=1$ (corresponding to the solution $a=a_0:\mathcal{L}_{\mathcal{N}}=0$ for  $r=0$ and $\sigma=1$). In general,  $\sigma\in]0,1]$, the bundle origin spin is $\la_0\equiv 1/\omega\sqrt{\sigma}$.

\medskip

\textbf{Replicas}

Replicas are light-like (circular) orbits, of  \textbf{NSs} as \textbf{BH}s spacetimes having the  same frequency  as a \textbf{BH}   $\omega_H^\pm$.
All  the points of a \MB\phantom{}    curve are the \emph{replica}   of the \textbf{BH} horizon frequency  $\omega_H^+$ or $\omega_H^-$ of the  \textbf{BH} individuated by the tangent point of a \MB\phantom{}    curve to the horizon curve  $(a_g,r_g)$.
The study of  replicas is the study of the \MB s curves in the extended plane.

Per definition all  orbits of a bundle are a replica of the horizon the bundle is tangent to.
More precisely, replicas   connect measurements in different    spacetimes  characterized by the same value of the property $\Qa$ function of $\omega$, by connecting   the  two null vectors, $ \mathcal{L}(r_\star,a_\star,\sigma_\star)$ and
$ \mathcal{L}(r,a,\sigma )$,  to $ \mathcal{L}(r_g,a_g,\sigma_g)$ where $r_g$ is the tangent radius of the \MB \  to the outer or inner Killing \textbf{BH} horizon  and $(\sigma_g,a_g)$,
$(\sigma_\star,a_\star)$ are generally different from $(\sigma,a)$.
In some cases,  horizons (frequencies) replicas are in the same spacetime ($a=a_\star=a_g$).

More precisely, the orbital frequency $\omega $ of a photon on a replica    coincides with the \textbf{BH}  horizon  frequency, which is also the  bundle characteristic frequency.
Although  here we restrict our analysis to the co-rotating replicas, i.e.  $\omega>0$ as $a>0$, the study of counter-rotating photon circular orbits with frequency $\omega<0$ is possible  with   \textbf{MB}s in the extended plane $a\in \mathrm{R}$ with $\omega>0$.
Notably not all \textbf{BH}  frequencies are replicated.
There is a \emph{confinement}, when   a horizon frequency   is not replicated i.e. there is no replica connected to that frequency.

On the equatorial plane an exact expression of the Kerr inner  horizons replicas  in a fixed \textbf{BH}  spacetime $a$,
$r^{-}_{-}:\quad  \omega_-(r_-^-)=\omega_-(r_-)=\omega_H^- $ and outer horizons replicas
$ r_+^+:\quad\omega_+(r_+^+)=\omega_+(r_+)=\omega_H^+$ are
\bea\label{Eq:mart-re}
&&
r^{\mp}_{\mp}=\frac{1}{2} \left(\sqrt{\frac{32 r_\mp}{a^2}- a^2\pm 6 \sqrt{1-a^2}-22}-r_\mp\right).
\eea
Since
 $0< r_-^-<r_-<r_+<r_+^+$ -- see Figs\il(\ref{Fig:PlotBri02m},\ref{Fig:PlotBri02mnsz},\ref{Fig:JirkPlottGerm1})-- while the outer horizon replicas can be detected on the equatorial plane, the inner horizon replicas remain  confined to the observer.

In the  Kerr spacetime, part of the inner horizon frequencies are  ``confined".
However, importantly for the observational point of view,    some replicas of the  inner horizons  are in the proximity of the \textbf{BH} poles, while  confined on different  planes  $(\sigma<1)$.
In \cite{nuclear,GRG-letter},   the  \textbf{BH} poles $(\sigma\approx 0)$ have been  studied  for the observation of the photon orbits with \MB s characteristic frequency. One of the intriguing   applications of \MB s in \textbf{BH} physics is the possibility to explore the regions close to the \textbf{BH} rotational axes (poles), where  it is possible to measure  the Kerr inner horizons replicas--\cite{nuclear,GRG-letter}.
In Figs\il(\ref{Fig:PlotBri02m},\ref{Fig:PlotBri02mnsz}), we show the   \textbf{BH}  horizons  replicas in some  \textbf{NSs} spacetimes of the two bundles of the Kerr extended plane $\la_0-r$ (at different $\sigma$), defined by the characteristic frequencies $\omega_H^\pm$, respectively, for the spacetime with $a=0.99M$, hence with tangent  equal to the tangent spin $a_g=0.99M$, but different tangent radius $r_g=r_\mp$, corresponding to the inner and outer  horizon of the \textbf{BH}  with spin $a=0.99M$.

Therefore, replicas in a fixed geometry,  allow  an observer to detect the horizon frequency on a different orbit. The topology of the \MB s curves in the extended plane   provides  information about  the  local properties of the spacetime replicated in regions  more accessible  to observes, which  can  detect  the presence of a replica at the point $(r_\star,\sigma_\star)$ of the \textbf{BH} spacetime with spin $a_\star$, by measuring the \textbf{BH} horizon frequency   $\omega_H^+(a_\star)$ or $\omega_H^-(a_\star)$  (the outer or inner \textbf{BH} horizon frequency)  at the point $(r_\star,\sigma_\star)$.

 The stationary observers  orbital  frequency  is  $\omega\in]\omega_-,\omega_+[$ where one of  ($\omega_{-},\omega_{+}$) is the horizon's frequency $\omega_H^{+}$, replicated on a pair of orbits $(r_+,r_s^*)$, where $r_s^*$ is  a light surface orbit, i.e. $r_s^*=r_s^+$ or $r_s^*=r_s^-$.     The second  light-like frequency  is the frequency of a horizon in a \textbf{BH}  spacetime.

From Figs\il(\ref{Fig:PlotBri02mnsz}) is it possible to see that, for slow spinning \textbf{NSs}, the inner horizons co-rotating replicas  change slowly, increasing the singularity spin-mass ratio.  The outermost co-rotating replicas of the outer horizon  change slowly,  increasing the \textbf{NSs} spin with respect to the  outermost horizon replicas.  The innermost outer horizon replicas change from very slow spinning \textbf{NSs} (upper right panel), slow spinning \textbf{NSs} and very fast spinning \textbf{NSs}. It is clear the change of the \textbf{NSs}  replicas between in \textbf{the} spacetimes of very slow spinning \textbf{NSs} (which is similar to the  case of  \textbf{BH}  spacetimes), slow spinning \textbf{NSs} and faster spinning \textbf{NSs}.  For very slow spinning \textbf{NSs} the replicas are very close to the inner and outer horizons  (upper right and left panels), describing the \textbf{NSs} bottleneck and horizon remnants.

\begin{description}

\item[\textbf{Horizons curve in the extended plane}]

The horizon curve $a_{\pm}$, solution of $r_\pm=r$, boundary of the \textbf{BH}  region in the extended plane,  is
\be\label{Eq:horiz-curve}
a_{\pm}\equiv\sharp\sqrt{r(2-r)}, \quad (\mbox{where}\quad r\in[0,2])
\ee
where  $\sharp = \pm$,  for the co-rotating $(\omega>0)$ and counter-rotating $(\omega<0)$ orbits, respectively.

\item[\textbf{Frequencies in the extended plane}]

Explicitly, the  limiting frequencies  {$\omega_{\pm}: \mathcal{L_N}=g(\mathbf{\mathcal{L}, \mathcal{L}})=0 $} are
\bea&&\label{Eq:bab-lov-what}
\omega_{\pm}\equiv\frac{\pm\sqrt{\sigma  \Delta \Sigma^2}-2 a r \sigma }{\sigma  \left[a^2 \sigma  \Delta-\left(a^2+r^2\right)^2\right]}, \quad  \mbox{where}\quad
 \lim_{r\rightarrow\infty}\omega_{\pm}=0\\	
 &&\nonumber \mbox{and}\quad
 \lim_{r\rightarrow 0}\omega_{\pm}=\frac{1}{a \sqrt{\sigma}}=\frac{1}{\la_0}\equiv \omega_0,
 \eea
 where we introduced the frequency function
 $ \omega_0$ expressing the bundle characteristic frequency as frequency at the origin $r=0$.
 Considering the limiting cases of the static   Schwarzschild  spacetime and of the faster spinning \textbf{NSs} geometry, we obtain
 \bea
&&\lim_{a\rightarrow0}\omega_{\pm}=
\mp\frac{(r-2)}{\sqrt{(r-2) r^3 \sigma }}\quad\mbox{and}\quad \lim_{a\rightarrow\infty}\omega_{\pm}=0.
\eea
We can write  the frequencies in Eq.\il(\ref{Eq:bab-lov-what}) as frequencies of the horizon curves in the extended plane as
\bea
&&
\omega_H^\pm(r)\equiv\omega_{\pm}(a_{\pm})=
 \frac{\sqrt{(2-r) r}}{2 r}\quad\mbox{where},\\
 &&\nonumber  r\in[0,2]\quad\mbox{and}\quad \lim_{a\to 0}\omega_H^\pm=0,
\eea
where $r\in[0,2]$ is  the horizons radius in the extended plane. Note that  $\omega_H^\pm=0$ for $r=2$.
The characteristic frequency $\omega$ is related to  the bundles origin $a_0\in[0,+\infty]$ (or $\la_0\in[0,+\infty]$),  and  the  radius $r\in[0,2]$  on the horizon curve in the extended plane   as follows
\bea&&\label{Eq:la0-origin}
\omega=\frac{1}{\la_0}=\frac{1}{a_0\sqrt{\sigma}}= \frac{\sqrt{(2-r) r}}{2 r}= \omega_H^{\pm}(r) \quad (\mbox{where}\quad r\in[0,2]).
\eea
Then,  $\omega_{\pm}=1/\sqrt{\sigma}$ for  the  extreme \textbf{BH}, and $\omega_{\pm}$  has a  minimum on the equatorial plane, where it is  $\omega_{\pm}=\omega_0=M/a$.

Note that  the
light surfaces $r_{s}^{\pm}$ on the Kerr spacetimes equatorial plane can be written as functions of the  frequency $\omega$ and parametrized in terms of the bundle characteristic frequency of the origin point $\omega_0$ as follows
\bea\label{Eq:rspm}
&&r_s^- \equiv \frac{2 \beta_1 \sin \left[\frac{1}{3} \arcsin\beta_0\right]}{\sqrt{3}}
,\quad
r_s^+ \equiv \frac{2 \beta_1 \cos \left[\frac{1}{3}\arccos(-\beta_0)\right]}{\sqrt{3}}, \\
&&\nonumber\mbox{where}\quad
\beta_1\equiv\sqrt{\frac{1}{\omega ^2}-\frac{1}{\omega_0^2}},\quad \beta_0\equiv\frac{3 \sqrt{3} \beta_1 \omega ^2}{\left(\frac{\omega }{\omega_0}+1\right)^2}
\eea
--see Figs\il(\ref{Fig:Plot-specif})--{lower} panel.

\item[\textbf{\MB s curve origin spin}]

\

Conversely,   the origin spin $a_0$ can be  read as
 \bea\label{Eq:def-a0}
 &&
  a_0=\frac{2}{\sqrt{\sigma}}\sqrt{\frac{r}{2-r}} =\pm \frac{1}{\omega\sqrt{\sigma }}\quad (\mbox{where}\quad r\in[0,2]). 
\eea

\item[\textbf{\MB s curve tangent radius}]

\

The \MB{}  tangent radius $r_g\in[0,2]$ to the horizons curve  can be expressed in terms of the frequency $\omega$ and the bundle origin spin  as
\bea\label{Eq:def-rg}
 r_g(\omega)=\frac{2}{4 \omega ^2+1}=\frac{2 \la_0^2 }{\la_0^2 +4}
\eea
--Figs\il(\ref{Fig:PlotClipMirr3}).

\item[\textbf{\MB s curve tangent spin}]

\

The \MB s  tangent spin   reads
\bea\label{Eq:def-ag}
 a_g(a_0)=4 \sqrt{\frac{\omega^2}{\left(4 \omega^2+1\right)^2}} =\frac{4 \la_0 }{\la_0^2 +4},
 \eea
--Figs\il(\ref{Fig:PlotClipMirr3}).
  \begin{figure*}
\centering
       \includegraphics[width=8.5cm]{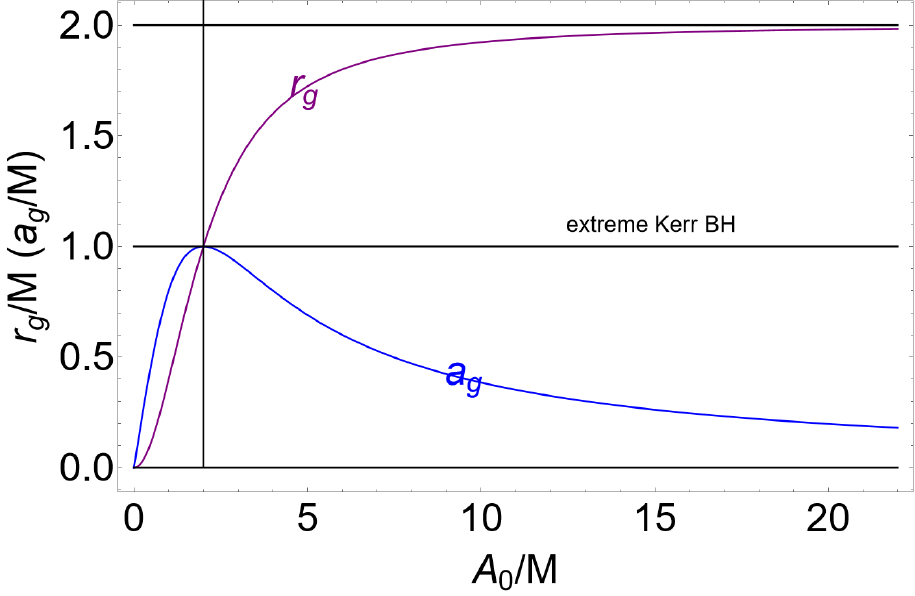}
        \includegraphics[width=8.5cm]{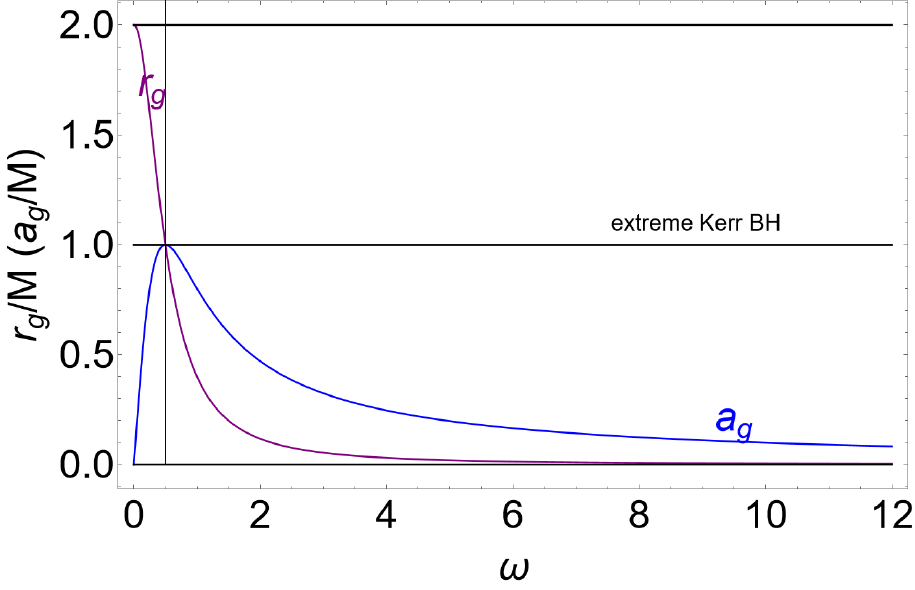}
           \caption{\MB s tangent radius $r_g$ (purple) and  tangent spin $a_g$ (blue) as function of the bundle origin spin $\la_0\equiv a_0\sqrt{\sigma}$. Here we show $\sigma\equiv\sin^2\theta\in[0,1]$) (upper panel) and the bundle characteristic frequency $\omega$ ({lower} panel). See discussion of Sec.\il(\ref{Sec:MBScharacteristics}). The radius $r_g$ is defined in Eq.\il(\ref{Eq:def-rg}) and tangent spin $a_g$ is defined in Eq.\il(\ref{Eq:def-ag}). The extreme Kerr \textbf{BH}  case, for $a_g=M$,  $r_g=M$, $\omega=1/2$, $\la_0=2M$ is also shown.}\label{Fig:PlotClipMirr3}
\end{figure*}
\end{description}

 In the next section we shall analyze  \MB s characteristics  for the limits of the \textbf{KN} solution as seen in the extended plane.

\subsection{Limiting cases}\label{Sec:limit-acses}

It is possible to  explain some \MB s properties   and their representation in the extended plane by considering some limiting cases, in particular the zeros of the \MB s curves in the extended plane.
Here, we discuss the limiting cases of the  static \textbf{RN} solution and electrically neutral and static limit of the Schwarzschild spacetime.

The definition of  the extended plane in the  Schwarzschild  spacetime is complicated  due to the identification of the    metric parameter  defining   \MB s.
On the other hand,  \MB s in the Schwarzschild case  can be studied as  zero limits  of the extended planes of the \textbf{RN} and Kerr spacetimes  (in the Schwarzschild case, the horizon is the axis $r=2M$ on the Kerr extended plane shown in Figs\il(\ref{Fig:Plot-specif})--left panel.).
More generally in static geometries,  the  bundle characteristic frequencies, solutions  of $\laa_{\mathcal{N}}=0$, are null and  the tangency condition of the metric bundles with respect to the  horizons curves is to be  understood as an asymptotic condition. In this situation,
   the horizon replicas are the  asymptotic solutions  for orbits  far from the gravitational source.

Below, first we analyze   Schwarzschild \MB s as the zeros of  the extended planes. Then,  we shall discuss  the
Schwarzschild \MB s as limits of the \textbf{RN} \MB s, concluding this section  with the analysis of the
Schwarzschild \MB s as limits of the Kerr \MB s.

\medskip

\begin{description}
\item[\textbf{The
Schwarzschild \MB s as zeros in the extended planes}]

\

The  Schwarzschild metric  case  corresponds to the zeros  of the metric bundles of the Kerr and \textbf{RN} spacetimes  (solutions with  $a=0$ and $Q=0$, respectively).
The Schwarzschild limiting photon orbital  frequencies are
\bea&&\label{Eq:rLmrLp}
\omega=\pm\frac{\sqrt{r-2}}{r^{3/2} \sqrt{\sigma }},
\eea
where    $\omega=0$ on the \textbf{BH} horizon ($r_+=2$).

\item[\textbf{The
Schwarzschild \MB s as limits of RN \MB s}]

\

Let first discuss the \textbf{RN} case, where the horizons are represented as the curve  $Q_\pm=\sqrt{r(2-r)}$ in  the  extended plane $Q-r$.
Let us consider   the Killing field $\laa$ and the quantity $\laa_{\mathcal{N}}$:
\bea
&&
\laa^a\equiv \xi_{(t)}^a+\omega\xi_{(\phi)}^a\quad  \laa_{\mathcal{N}}\equiv{g}(\laa,\laa)=g_{tt}+g_{\phi\phi}\omega^2.
\eea
From the equation $\laa_{\mathcal{N}}=0$,
we find the limiting (photon orbital) frequencies $\omega_\pm$ for the \textbf{RN} geometry:
\bea\label{Eq:be-en}
&& \omega_{\pm}=\pm\frac{-g_{tt}}{g_{\phi\phi}}={\pm} \frac{\sqrt{Q^2+(r-2) r}}{\sqrt{\sigma}r^2}
\eea
where  $\omega_\pm=0$ on $r_\pm$.
{Differently from  the Kerr case given in Eq.\il(\ref{Eq:bab-lov-what}), the  frequencies functions in Eq.\il(\ref{Eq:be-en})} are even in the electric charge parameter $Q$. In a static spacetime, we can define relatively counter-rotating photon orbits and following the same procedure in the spherically symmetric background, the expression of the \MB s in the \textbf{RN} spacetime reflects this property.
 \textbf{RN} \MB s in the  extended plane can be represented as the function
 \bea
Q(r)\equiv\sqrt{r} \sqrt{r^3 \sigma  \omega ^2-r+2}.
 \eea

\item[{\textbf{The Schwarzschild \MB s as limits of the Kerr \MB s}}]

Metric bundles of the  Schwarzschild geometry  are  the limiting case of the  metric bundles of the Kerr geometry for $a=0$. In this case, $\omega=0$ on the horizon.
From  Eqs\il(\ref{Eq:def-rg}), (\ref{Eq:def-a0})
and (\ref{Eq:def-ag}), we obtain
 \bea\label{Eqs:limits-Kerr-to-static}
&&\lim_{\omega\to0}r_g(\omega)=2,\quad  \lim_{\omega\to 0} a_0=\infty
\quad \mbox{and}\quad \lim_{a_0\to\infty}a_g=0.
\eea
The first and second limits define  the static case for the  null horizon frequency $\omega=0$,  described as a limiting situation occurring for the horizon tangent point $r_g=2$ (the Schwarzschild horizon radius), corresponding to the zero of the (outer) horizon curves  in the Kerr extended plane. The frequency $\omega=0$
corresponds also to the \MB s origin spin $a_0$
for extremely fast spinning \textbf{NSs}. The \MB s  in the limiting case of very slowly spinning \textbf{BH}s  have origins corresponding to extreme fast spinning \textbf{NSs}, as it is clear from the
 third limit.
 \end{description}

\section{Black hole thermodynamics} \label{Sec:embo}
Israel's theorem  concerns  the uniqueness of the Schwarzschild \textbf{BH}  (the only static and asymptotically flat
vacuum spacetime with  a regular horizon is the Schwarzschild solution) and  establishes that the event horizon must be spherically symmetric, implying  that  the sphericity of
an isolated and  static \textbf{BH} cannot be broken \cite{Israel1,Israel2} and \cite{HE,H72152}.
This is  the first version
of the so called   ``no-hair theorem", later   formulated  by Wheeler, and  which today  refers more generally to a set of theorems and conjectures.
The first result of Israel  established
the uniqueness of the Schwarzschild metric, which was then extended to the case of spinning  and charged  \textbf{BH}s \cite{Carterspinning,Robinson}.
Although some steps have been made towards a full generalization of the Israel theorem, there is still no  proof for a general no-hair theorem remaining, therefore, a conjecture.

 The   ``no--hair theorem"  states  also what is known as the simplicity of the  isolated \textbf{BH}s in equilibrium,  which  (in General Relativity) can be fully characterized by the mass $M$,
the angular momentum $J$, and the electric charge $Q$.
Said differently,  all stationary \textbf{BH}  solutions of the Einstein--Maxwell equations are defined by only three independent  (classical and observable) parameters.
The term  ``hairs" refers to any further   ``information" as,  for example,  the matter of the \textbf{BH}  progenitor or  other  ``conserved numbers" of the matter swallowed by the \textbf{BH}, which  could not be traced back by a far away observer. Hence,  the physical  \textbf{BH}   (as final result of a physical process in a progenitor lifetime) after collapse has left with no  other  defining (independent)  ``remnant" parameters, when  the matter-field information (prior horizon formation /crossing)     ``disappeared" behind the \textbf{BH}   horizon, inaccessible to the  outside observer and, therefore, lost to the observer's universe.

 Hairs, in other words, are fields  (different from the electromagnetic ones) and matter,  which is associated to and  determines stationary \textbf{BHs}.

The conjecture   also outlines  the  \textbf{BH}   information paradox  (problem) in its   classical  and   quantum formulation,  when considering the possible quantum effects associated to the \textbf{BH}  and the spacetime close to the horizons.
This and the approximation of the general formulation of the conjecture has led to several reformulations, revisions, and modifications to the initial conjecture, as confirmations or exceptions were encountered in relation to specific  ``hairs--fields" \cite{soft1,soft2,soft3}.


In the first theorem formulation, the electromagnetic  field was singled out (excluding, however, a magnetic charge), and excluding further  ``degrees of freedom" as the  Baryon number,  for example, or other quantum numbers  (as  strangeness,  etc),   which, therefore, could not be measured by a (far away) observer, being destroyed   during the collapse (or crossing the horizon).

Therefore, the theorem had the further issue to isolate,  without physical  foundation,  the electromagnetic  field  (gauge fields subjected to Gaussian law ) with respect to other fields. Several counter--examples have been found with fields hairs for stationary and stable \textbf{BH}  solutions.
There are   stationary \textbf{BH}s  with  hairs as
exterior fields  constituting exceptions.
 Stable stationary \textbf{BH}s  solutions with gauge  field hair
are also known (the \textbf{BH}  pierced by   cosmic strings are
another example of hair as  other solutions in
non-abelian gauge theories.).

Remarkably, there is no clear, univocal  and physically supported indication (or criterion) on the   fields to consider as possible \textbf{BH} -- ``hairs",  this constituting  a deeper and broader issue  in the  ``no-hairs" theorem investigation.
It should be noted that mass, electric charge and angular momentum  are  all conserved quantities  holding the Gauss law. An eventual  magnetic charge, according to this requirement, would be a  ``proper " \textbf{BH}  hair.
However,    many \textbf{BH}   solutions with gauge fields (different from  the Maxwell fields),  subjected to a  Gauss law were also found.

Therefore,  the formulation  of a no-hair theorem itself  is under scrutiny, so that more than one theorem is to be proved; it resulted into  an idea to be tested on a case-by-case basis which, with negative results, confirms the conjecture for that specific case, or provides a counter-examples by narrowing the field of analysis on a possible geometric justification of the \textbf{BH}  hair fields, giving raise to a sequence of  ``no--hair
theorems" (there are, for example, non scalar--hair theorem, etc).
It would, therefore, be more correct to say that,
in the context  of General Relativity, the no-hair theorem refers to a series of  statements concerning the  ``degrees of freedom" defining  a \textbf{BH}  and  observable to a distant observer.
Another issue of the no-hair theorem is the  role of  conditions at infinity (and the asymptotic  flatness),  relevant, for example, in cosmological models, where formulations are known for positive cosmological constant--see \cite{BaPRL}.
 Obviously a
no-hair theorem would imply the \textbf{BH}   ``simplicity" and would  have consequences on the \textbf{BH}   uniqueness.
A conceptual consequence of the no--hair theorems (\textbf{BH} simplicity) is  that in  this way the \textbf{BH}   entropy,  defined by the \textbf{BH}   area,  would provide   the  ``real" measure of the  \textbf{BH}   degrees of freedom.
 However,  the information  on the \textbf{BH}  ``past" (with respect to the state of observation),  in the sense of  \textbf{BH}  ``hair", is to be considered eliminated or
  inaccessible  to  far away  observers.

This field of investigation with its problematic aspects is  a promising cross-road among \textbf{BH}  physics, quantum gravity, information theory, and theories of great unification. Nevertheless,  this relevant aspect of \textbf{BH}  physics   diverges from the purposes of this chapter and   we refer the reader, for   further details on this issue, to the extensive literature on the subject.


Here
 we explore the laws of \textbf{BH} thermodynamics    in the extended plane,
using the Smarr's formula  connecting, in the Kerr \textbf{BH}  spacetime,  $(M,J,A_{area})$, where $A_{area}$ is the (outer) \textbf{BH}  horizon area \cite{Smarr}.
The extended plane  and   \textbf{MBs}  can be  used  to relate  different states of a  \textbf{BH}, interacting  with its  environment.
Replicas can  connect \textbf{BH} spacetimes related in a transition and governed by the thermodynamic laws   in the extended plane. Hence, it is possible  to reformulate  \textbf{BH} thermodynamics  in terms of the light surfaces, exploring the thermodynamic properties of the geometries through the metric bundles.

 This formalism turns out to be adapted to the description of the \textbf{BH} state transitions,  constrained  by the properties of the null vector  $\mathcal{L}$,  building up many of the thermodynamic properties of \textbf{BHs}, as it enters the definitions of thermodynamic variables and stationary observers.

In this  section, we  detail the \textbf{BH} thermodynamic properties in terms of the null  vector $\mathcal{L}$ and, therefore, of  the metric bundles.
 Sec.\il(\ref{Sec:disc-MBL})  discusses the role of  metric bundles and the quantity
$\mathcal{L_{\mathcal{N}}}$ in \textbf{BH}  thermodynamics,  introducing  the concept of  \textbf{BH}  surface gravity, the first law of \textbf{BH}  thermodynamics, and  concluding with some notes on the \textbf{BH}s states transitions as described by the laws of \textbf{BH}  thermodynamics.

In this framework, we explore \textbf{BH}  states transition in the extended plane in Sec.\il(\ref{Sec:nil-base-egi}).
The  \textbf{BH} irreducible   mass is the subject of Sec.\il(\ref{Sec:sud-mirr-egi}), where
 we discuss also the
    extraction of \textbf{BH}   rotational energy as constrained  with \MB s.
 Finally, Sec.\il(\ref{Sec:fin-BH-THEr-presenT}), \textbf{BH}  thermodynamics in the extended plane, closes this section.

\subsection{Black hole thermodynamics, metric bundles and the quantity
$\mathcal{L_{\mathcal{N}}}$}\label{Sec:disc-MBL}

In this section,  we  discuss some concepts of  \textbf{BH} thermodynamics in terms of   metric bundles and the quantity
$\mathcal{L_{\mathcal{N}}}$. First, we introduce  the concept of  \textbf{BH}  surface gravity, we then  discuss the  first law of \textbf{BH}  thermodynamics,  concluding the section  with some notes on the \textbf{BH}s states transitions.

\medskip

\textbf{\textbf{BH} surface gravity }

We  start  our analysis of  the \textbf{BH} thermodynamics by introducing  the concept  of   \textbf{BH}  surface gravity and derive a surface gravity function in the extended plane, expressed  in terms of \MB s characteristics.

  The surface gravity $\ell$ for a \textbf{BH}  Killing horizon is defined by the fact that the Killing
vector defines a non-affinely parametrized geodesic on the Killing horizon (a global  Killing vector field $\mathcal{L}^a$
becomes null on the event horizon).
According to the  Zeroth Law of \textbf{BH}   thermodynamic,
the surface gravity  of a stationary  \textbf{BH}  is constant over the
event horizon (somehow fixing a concept of \textbf{BH}  thermal equilibrium)-- see for example \cite{WW}.

In general terms, given an object and its  surface, we can define the surface gravity  as the (gravitational) acceleration  a  test particle  in the proximity of the object is subject to. Then,
a \textbf{BH} surface gravity is  the  acceleration, as exerted  at infinity,  of a test particle located  in close proximity to  the \textbf{BH} outer horizon, which is necessary to keep it  at the horizon.
For a static \textbf{BH}, it  can be expressed  as the force exerted at infinity to hold a  test particle at close proximity  of  the horizon.
 Notice that  the locally exerted force  at the horizon is infinite.

 A similar definition holds for  stationary \textbf{BH}s and, in general, for   \textbf{BH}s with   ``well  defined"  Killing horizons.
However, surface gravity   is a classical geometrical concept, playing a major  role in \textbf{BH} thermodynamics--\cite{BCH}.
Bekenstein  first suggested  the analogy leading to the concepts of
 \textbf{BH} entropy and  temperature   and the  formulation  of  classical \textbf{BH} thermodynamics--\cite{Bekenstein73,Bekenstein75}.

 Hawking later  confirmed Bekenstein's conjecture of a reliability of a \textbf{BH}  entropy definition,  establishing  the (correct) constant of proportionality relating the \textbf{BH}  entropy and \textbf{BH}  area, following  a more complex and complete argument  relating  \textbf{BH}  thermodynamics and \textbf{BH}  energy extraction \cite{Hawking75,H77,SV96}.

The introduction of a temperature  leads naturally to inquire about a possible \textbf{BH}  radiation emission,  which would be regulated by the temperature defined by its surface gravity.
(However, although the \textbf{BH}  temperature could be null, the \textbf{BH}   entropy could not vanish).
An interpretation of this suggestion has been  realized  in the  quantum (semi-classical) frame.
 \textbf{BH}  thermal radiation has been
 shown  to be  the  result of  quantum mechanical effects governed by the \textbf{BH} temperature and, therefore, its surface gravity (Hawking semi-classical  radiation). The surface gravity regulates also the
 probability of a negative energy particle tunneling through the horizon--\cite{Hawking75}.
 Because it radiates, one can expect  the \textbf{BH} mass to decrease  and   eventually
disappear. Nevertheless,  the actual fate of the \textbf{BH} and the information carried in the  \textbf{BH} during and before  the evaporation process,  the possible existence of  remnant at the end of process, and the entropy evolution are matters of an  ongoing (vivacious) debate--see for example \cite{JP16,LS,Hawking05}.
The actual fate of a \textbf{BH}, following the characteristic thermal emission, is an  example of controversial aspects of the surface gravity physics.

Thus, the \textbf{BH}   (Bekenstein-Hawking) entropy links  gravitation, \textbf{BH}  thermodynamics, geometry, and quantum theory.
The so-called  (\textbf{BH}) information paradox  is seen  today as the source of one of  most promising hints for a  quantum  theory of gravity.

The relation between area and entropy and  the discussion around  their  constant of proportionality had  enormous relevance far beyond the study of \textbf{BH}s. The
  Bekenstein bound was  the  starting point,  for example, of the holographic principle and in general shred light on profound aspects of geometric theories  and problems of  quantum information theory.

The history of     Bekenstein bound confirmations, developments and applications is broad and complex. As it goes beyond the discussion in this chapter,  we refer   the interested reader to the extensive literature on the subject.
Here, we will reformulate the relation (proportionality) between the concept of entropy and \textbf{BH}  area,   in terms of light surfaces through  the \MB s.

\medskip

	 Formally,  surface gravity   may be   defined as the  {rate} at which the norm   of the Killing vector $\mathcal{L}$
vanishes from
outside (i.e. from $r>r_+$). In fact, $\ell$ is  (the constant) defined through the relation
$\nabla^a\mathcal{L_{N}}=-2\ell \mathcal{L}^a$ (on the outer horizon).

Equivalently,
  $\mathcal{L}^b\nabla_a \mathcal{L}_b=-\ell \mathcal{L}_a$ and  $L_{\mathcal{L}}\ell=0$, where $L_{\mathcal{L}}$ is the Lie derivative,-a non affine geodesic equation, i.e.,
$\ell=$constant on the orbits of $\mathcal{L}$.
(The norm {$\mathcal{L_{N}}\equiv g(\mathcal{L},\mathcal{L})$} is  constant on the horizon.)

 For
the Kerr spacetime it is $\ell= (r_+-r_-)/2(r_+^2+a^2)$.
In the extreme Kerr  spacetime ($a=M$), where  $r_{\pm}=M$, the surface gravity  is zero and the \textbf{BH}   temperature  is also null ($T_H = 0$), but  with a \emph{non-vanishing } entropy \cite{Wald:1999xu,WW}.
This relevant  fact establishes a profound  topological difference between \textbf{BH}s and extreme \textbf{BH}s, implying  that  a
\textbf{BH} cannot reach the   extremal limit in a finite number of steps--(third law of \textbf{BH}  thermodynamics)--  having  consequences
also regarding the stability
 against Hawking radiation.

On the other hand, the condition (constancy of )  $\nabla^a \laa=0$  when $\ell=0$ substantially constitutes the definition of the  degenerate Killing horizon \textbf{BH}. In the case of Kerr geometries, only the extreme \textbf{BH} case is degenerate. Therefore, in the extended plane it corresponds to the point ($a=M$, $r=M$).
(The \textbf{BH} surface gravity  $\ell$,  is also a  conformal invariant of the metric.
	Furthermore the surface gravity   re-scales with the conformal Killing vector, i.e. it  is not the same on all generators but,  because of the symmetries,  it is constant along one specific generator--see also \cite{J-S09,Jacobson:2010iu}.)

\medskip

\textbf{First law of \textbf{BH} thermodynamics}

\medskip

The   Kerr \textbf{BH} area is given by  the function
\bea&&
A_{area}=\int\limits_{0}^{\pi}d \theta\int\limits_{0}^{2\pi} \sqrt{g_{\phi\phi}g_{\theta\theta}}  d\phi=4 \pi  \left(a^2+r^2\right),
\eea
evaluated on the outer horizon $r=r_+$.
The \textbf{BH} (horizon) area $A_{area}^+$ is  related to the outer horizon definition,  $A_{area}^+ = 8\pi mr_+$, where  $M = c^2m/G =m$ in geometric units.
{(For simplicity, in some of the expressions  we do not consider the factor  $8\pi$ and in some expressions we write
  spin, radius, characteristic frequency and origin spin as dimensionless parameters.)}

The  \textbf{BH}  horizon area
is non-decreasing, a property which is considered as the second law of
\textbf{BH}  thermodynamics (establishing  the impossibility  to achieve
with any physical process a \textbf{BH} state with zero surface gravity).
Note that the (Hawking) temperature term is  related to the surface gravity by $T^+_{H}= {\hslash c\ell^+ }/{2\pi k_{B}}$, while the  horizon area $A_{area}^+$ defines the \textbf{BH}  entropy
as  $S^+= k_{B} A_{area}^+/l_P^2$, $l_P$ is the Planck length,  $\hslash$ the reduced Planck constant, and $c$ is the speed of light,  $k_{B}$ is the Boltzmann constant.
(We use the  notation $(+)$   or $\Qa_H^+$  for   all the quantities $\Qa$,  evaluated on the outer Killing horizon.)

The first law of  \textbf{BH} thermodynamics,
  $\delta M = (1/8\pi)\ell_H^+ \delta A^+_{area}+ \omega^+_H \delta J$, relates the
variation of the mass $\delta M$, to  the (outer) horizon area $\delta A^+_{area}$, and angular momentum $ \delta J$
with the surface gravity  $\ell_H^+$ and angular velocity $\omega_H^+$ on the outer horizon.
In this expression,
  the term  $p^+= - \omega_H^+ $, could be seen as    ``pressure-term",  where the
internal energy is $U= GM$. The term $\omega_H^+ \delta J$ is  interpreted as the ``work''.
 The volume term is $V= G  J/c^2$ (where $J = amc^3/G$).

 We shall consider   generalized definitions  in the extended plane, where we express some of the concepts of \textbf{BH} thermodynamics in terms  of the norm
$\mathcal{L}_{\mathcal{N}}$, which defines  the
metric bundles.
It will be necessary to consider quantities evaluated on the inner horizon, where we use the notation   $(-)$ or $\Qa_H^-$ for a quantity $\Qa$.

\medskip

\textbf{BHs transitions}

We can describe    the \textbf{BH} transition from an initial  state $(0)$ to a new state $(1)$ in terms of the characteristic bundle frequency (appearing explicitly in the first law  as  the work term  $\omega^+_H \delta J$).  We use the notation  $(0)$ and $(1)$ to  indicate any quantity related to the initial and final state, respectively; therefore,  $\omega_H^+(0)$  is the  \MB{} frequency  tangent to the outer horizon of the \textbf{BH} at  the initial state $(0)$.  The notation $\delta\Qa\equiv\Qa(1)-\Qa(0)$ denotes the change of the quantity $\Qa$ from  the initial $(0)$ to the final state $(1)$ of the transition. As in the bundles framework we evaluate  quantities in the extended plane, we express many quantities considered here as evaluated on the inner horizons.
There is a  relation  between the quantities  $(\delta A_{area}^-, \delta J, \delta M)$ and $(\omega_H^-(0),\ell^-(0))$ evaluated on the inner horizon $r_-$.
 All the quantities are expressed in terms of  bundles at $\omega_H^\pm(0)=$constant and $(\delta A_{area}^\pm, \delta J, \delta M)$, describing the transition from the initial to final  state.
\subsection{Masses and \textbf{BH} thermodynamics}\label{Sec:nil-base-egi}
We  proceed relating  the parameters
$(\ell^{\pm}, \omega^{\pm}_H)$, regulating the \textbf{BH} transition from one state to another, represented as a transition from a point $r(0)$ to a point  $r(1)$ along the horizons curve in the extended plane in the following  two cases: \textbf{\textit{1.}}
$(r(0),r(1))\in[0,1]$, inner horizon range, {and} \textbf{\textit{2.}} $(r(0),r(1))\in[1,2]$, outer horizon range (note,  we included the horizon  $r_\pm=1$ for the extreme Kerr spacetime  in both the inner and outer horizons  ranges).
We will   connect properties defined on   the  outer horizons curve of the extended plane  with  properties defined on the   inner horizons curve of the extended plane.

The frequency $\omega_H^\pm$  will  be considered  always  positive or null,  i.e.,  $J\omega_H^\pm\geq0$, or $a\omega_H^\pm\geq0$, vanishing  only for $J=a=0$. This implies  that we will not consider  counter-rotating orbits, nor we shall connect proprieties defined in the positive section of the extended plane to the negative section of the extended plane.
 However,  with $J\omega_H^\pm\geq0$, the surface gravity $\ell_H^\pm$ can be negative, when evaluated on a point of the inner horizon ($\ell_H^-\leq0$).

We consider the  quantity $s$   such that
 $\Pa_{\pm}(r(0))=s \Pa^*_{\pm}(r(1))$,  where $\Pa_{\pm}\in\{\ell_H^\pm,\omega_H^\pm\}$ and the notation $(*)$ represents an eventual change in sign.
{Note that   $\delta A_{area}^+=-\delta A_{area}^-$  for  $\delta M=0$ (as, per definition, $\delta M^+=\delta M^-\equiv\delta M$).}
For a fixed \textbf{BH}   spacetime
   $\delta M^+=\delta M^-$,  but the relation regulating this variation with the other characteristic  quantities depends on the point on the horizons curve.

More generally, we can consider   $\delta M^{\star}(a(0))=s  \delta M^{\ast}(a(1))$, where $(a(0),a(1))\in [0,M]$,  $s$ is a constant,
    and  for $a(0)=a(1)$  there is  $(\star=\ast=\pm, s=1)$.
In general, it  can be proved that
\bea&& \delta M^+=\ell_H^+\delta A_{area}^++\omega_H^+\delta J^+=\delta M=\\
&&\nonumber\ell_H^+\delta A_{area}^++\omega_H^+\delta J =
\delta M^-\equiv
\ell_H^-\delta A_{area}^-+\omega_H^-\delta J^-.
%
\eea

Assuming  $s\geq0$ and  $s\neq 1$, we  consider the relations   $
 \omega_H^*(r(0))= s\omega^\star_H(r(1))
  $ and $
 \ell_H^{\circ}(r(0))=- s \ell^{\diamond}_H(r(1)) $, where $(\circ,\diamond)=\pm$.
Note that s
$ \ell_H^+(a)\geq 0$  (surface gravity as function of the spin $a$ in the extended plane) and $\ell_H^-(a)\leq 0$, where
$ \ell_H^+(a)<-\ell_H^-(a)$. Also,
$\ell_H^\circ(s)<0$ for $s\in]0,1[$ and
$\ell(r)\gtreqless
 0$ for  $r\gtreqless 1$ (surface gravity/acceleration as function of the radius $r$ on the extended plane).
We obtain a horizon parametrization in terms of the constant $s$,   where $r(0)$ and $ r(1)$  correspond to the  horizons of the \textbf{BH} spacetime  with  $a(r(0))=a(r(1))$.  However, the limiting case $s=1$ corresponds to the extreme Kerr \textbf{BH}. Also, for $s\in]0,1[$ we have that  $r(1)=r_-(1)$ and $r(0)=r_+(0)$,  and  for $s>1$, it holds
 $r(1)=r_+(1)$ and $r(0)=r_-(0)$, where
 \bea\label{Eq:rrp}&&
 \frac{r(1)}{r(0)}=s,\quad
 r(1)=\frac{2 s}{s+1},\quad
\omega(r(1))=\frac{1}{2 \sqrt{s}},
\quad  \ell(r(1))=\frac{s-1}{4 s}\\&&\nonumber\mbox{where}\quad a(r(1))=a(r(0))=2\sqrt{\frac{s}{(s+1)^2}}
\eea
($\omega$  is the bundle characteristic frequency).
It is noteworthy that the $s$ factor is different according with the spin, i.e., expressed in terms of the spin it is  $s=\hat{s}_{\mp}\equiv[({2-a^2})\mp 2 \sqrt{({1-a^2})}]/{a^2}$.

\subsection{The  \textbf{BH} irreducible   mass}\label{Sec:sud-mirr-egi}
The total  \textbf{BH} mass $M$ of the Kerr \textbf{BH}
 can be decomposed into the  mass $M_{irr}$ and  the
 rotational energy.
From the first law of \textbf{BH}  thermodynamics, we obtain
  $M^2= M_{irr}^2+J^2/4M_{irr}^2$  (note, $J$ has   units   of mass $M^2$ and it is  measured in the asymptotic flat region).
The Kerr  \textbf{BH} irreducible (or rest) mass $M_{irr}=\sqrt{a^2+r_+^2}/2$, where $A_{area}=16 \pi M_{irr}^2$ (evaluated on the outer horizon). It  follows that
%
%
\bea\label{Eq:mass-irr-rampi}&&
\frac{M^+_{irr}}{M}=
+\frac{\sqrt{1+\sqrt{1-\frac{J(0)^2}{M(0)^4}}}}{\sqrt{2}},
\eea
or, more generally {the four solutions
\bea\label{Eq:seven-t0}
 \frac{M_{irr}}{M}=
\flat\frac{\sqrt{1\natural\sqrt{1-(a_{(\pm)})^2}}}{\sqrt{2}},\quad\mbox{where}\quad
\flat=\mp; \; \natural=\pm.
\eea}
Here,
$a_{(\pm)}$ is the dimensionless spin on the horizons curve, where
\bea\label{Eq:seven-t}
\frac{M^+_{irr}}{M^-_{irr}}=\sqrt{\frac{r_+}{r_-}}=\sqrt{\frac{\omega_H^-}{\omega_H^+}}=
\sqrt{\frac{-\ell_H^-}{\ell_H^+}}=\frac{\sqrt{r_+}}{\sqrt{2-r_+}}=\sqrt{\frac{1}{4(\omega_H^+)^2}}=\frac{\la_0}{2},
\eea
in terms of the bundles origin spin $\la_0\geq 2$. {(The notation $\pm$ refers to the radii $r_{\pm}$.)}
(Note that  $M_{irr}/M=\{0,1\}$
for $a=0$ and
 $M_{irr}/M=\pm1/\sqrt{2}$ for $a=M$.).

Equations (\ref{Eq:seven-t}) and (\ref{Eq:seven-t0})  are  in terms of quantities defined on the extended plane where, in terms of the  irreducible mass $M_{irr}$, we obtain
\bea&&\label{Eq:bullionss}
\la_0(all)= \frac{2 M_{irr}}{\sqrt{1-M_{irr}^2}}.
\eea
Here and in the following analysis, for
$\Qa(all)$  it is intended a generic  quantity $\Qa$, considering the entire horizon  curve in the positive section of the extended plane.

\begin{figure*}
\centering
                \includegraphics[width=8.5cm]{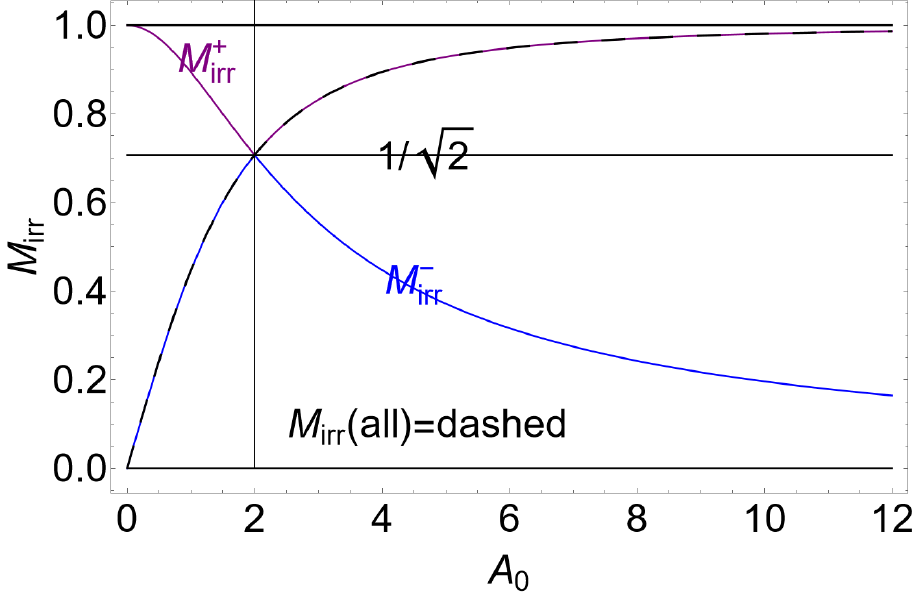}
           \caption{Irreducible mass function $M_{irr}$ versus bundle origin spin $\la_0\equiv a_0 \sqrt{\sigma}$ ($\sigma\equiv \sin^2\theta$). The functions $M_{irr}^\pm$ and
           $M_{irr}(all) $ are defined in Eq.\il(\ref{Eq:Mirr-all}). The limiting value $M_{irr}^\pm=M_{irr}(all)=1/\sqrt{2}$ for $\la_0=2$ (corresponding to the extreme Kerr \textbf{BH} ) is also shown. For $\la_0>2$ ($\la_0<2$) metric bundles are tangent to the outer (inner) horizons curves in the extended plane. See the analysis of Sec.\il(\ref{Sec:sud-mirr-egi}). All quantities are dimensionless.  }\label{Fig:PlotClipMirr2}
\end{figure*}

 Therefore, the \textbf{BH}  irreducible mass can be written as follows:
\bea &&\label{Eq:Mirr-all}M_{irr}(all)=\sqrt{\frac{\la_0^2}{\la_0^2+4}}=\sqrt{\frac{r}{2}}
=\sqrt{\frac{1}{4 \omega ^2+1}},
\\&&\nonumber\mbox{and}\quad  M_{irr}^\mp=\sqrt{\frac{1}{2}\mp\frac{1}{2} \sqrt{\frac{\left(\la_0^2-4\right)^2}{\left(\la_0^2+4\right)^2}}}=\sqrt{\frac{r_\mp}{2}},
\eea
%
--Fig.\il(\ref{Fig:PlotClipMirr2})--
($r$ is a  point on the horizon curve in the extended plane).

The dependence on $\sigma$, implicit in the   bundle origin definition $\la_0\equiv a\sqrt{\sigma}$,  does not contradict  the \textbf{BH}  horizon rigidity, as it has to be considered as a representation of the light surfaces frequencies on different poloidal angle $\sigma$s. In Eq.\il(\ref{Eq:Mirr-all}), we note the special  value $\la_0=2$, corresponding to a bundle characteristic frequency $\omega=1/2$, tangent to the horizon curve point, corresponding  to the extreme Kerr \textbf{BH}  $(r=r_\pm=1)$, where
$M_{irr}=1/\sqrt{2}$.

In terms of the bundle origin spin, we obtain the limiting values,
\bea
\label{Eq:o.r-childm}
&&\lim_{\la_0\to\mathbf{X}} M_{irr}(all)=\left\{1,0,\frac{1}{\sqrt{5}},\frac{1}{\sqrt{2}}\right\},\quad \mbox{and}\\
&&\nonumber \lim_{\la_0\to\mathbf{X}} M_{irr}(r_+)=\left\{1,1,\frac{2}{\sqrt{5}},\frac{1}{\sqrt{2}}\right\}
\eea
 for $ \mathbf{X}=\{+\infty,0,1,2\}$, respectively.

\medskip

\textbf{The  rotational energy $\xi$}

The  upper
limit of the   the amount of  energy  that can be extracted   from a Kerr \textbf{BH} is   the  \emph{total} rotational energy, occurring when  the Kerr \textbf{BH} final state is   a Schwarzschild \textbf{BH}  \cite{Ruffini}. The total mass  $M$  at the end of a (stationary)  process of the energy extraction has to be with  $M_{irr}$, which  is
   the  maximum  rotational energy that
can be extracted from the black hole is   $\xi\propto(M-M_{irr})$.
Hence, the maximum rotational energy which can be extracted  is the limit of $\xi=\xi_{\ell}\equiv \left(2-\sqrt{2}\right)/2$, where at the state $(0)$ (prior to the extraction) there is an extreme Kerr spacetime (with spin $a=M$).
In general,     $\xi\in[0,\xi_{\ell}]$.

The  extracted rotational energy $\xi$  is
 \bea\nonumber&&
\frac{\xi}{M(0)}=1-\frac{\sqrt{1+\sqrt{1-\frac{J(0)^2}{M(0)^4}}}}{\sqrt{2}}.
\eea
In the extended plane, we can express   the tangent (dimensionless) spin as function of the rotational energy parameter $\xi$:
\bea
a_{\xi}(\xi)\equiv 2 \sqrt{(2-\xi) (\xi -1)^2 \xi},
\eea
linking  the former state spin $a(0)$ to the rotational extraction in the subsequent phase,  where the \textbf{BH} is settled in a Schwarzschild spacetime.
More generally, solving   $a_{\xi}(\xi)\equiv a_{(\pm)}$ for $\xi$, we obtain {the four solutions
\bea
&&
\label{Eq:xis-nilo}\xi_s^{(\pm)}=1\flat\frac{\sqrt{1\natural\sqrt{1-a_{(\pm)}^2}}}{\sqrt{2}},\quad\mbox{where}\quad \flat=\mp; \quad \natural=\pm,
\eea }
where
$a_{(\pm)}$ is the dimensionless spin on the horizons curve.

The dimensionless spin  $a/M\equiv J(0)/M(0)^2$ refers to an initial state before the transition, defined by the mass  $M(0)$  and spin  $J(0)$, function of the extracted rotational energy  $\xi/M(0)=1-M_{irr}(0)/M(0)$, and, therefore, of the ratio  $M_{irr}(0)/M(0)$.
Measuring $\xi$, therefore, will provide and indication of the \textbf{BH} spin\cite{Daly0,Daly2,Daly3,
GarofaloEvans,ella-correlation},
 relating the  dimensionless \textbf{BH} spin $a/M$ to   the dimensionless ratio $\xi$ (as total released rotational  energy   versus \textbf{BH}  mass, measured by an observer at infinity, and assuming a  process ending with  the  total extraction of  the    rotational energy of the central Kerr \textbf{BH}).

We have that  $a_{\xi}(\xi)\in[0,M]$ (dimensionless)  with  $\xi\in[0,\xi_{\ell}]$, limiting, therefore, the (rotational) energy extracted (through a classical process) to a maximum of  $\approx 29\%$ of the mass $M$.
More in details:   $\xi=1$ for   $a_{\xi}({\xi})=0$   (the static limit corresponds also to the values   $\xi=\{0,2\}$). For  the extreme case, where  $a =
 M $ and  the frequency is $\omega= 1/2$, it is    $ \xi=1\pm {1}/{\sqrt {2}}$.
There is a maximum $\xi_{\ell}\equiv\left(2-\sqrt{2}\right)/2$ (and  $\xi_m\equiv \left(2+\sqrt{2}\right)/2$), where $a_{\xi}(\xi)=M$ and  $r_+=M$ (the extreme Kerr \textbf{BH}).

\begin{figure*}
\centering
       \includegraphics[width=8.5cm]{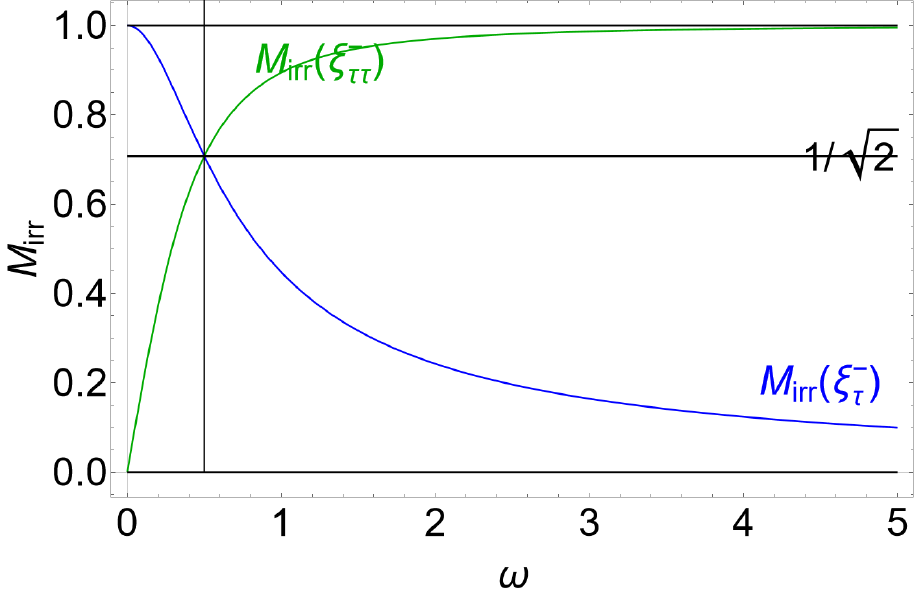}
        \includegraphics[width=8.5cm]{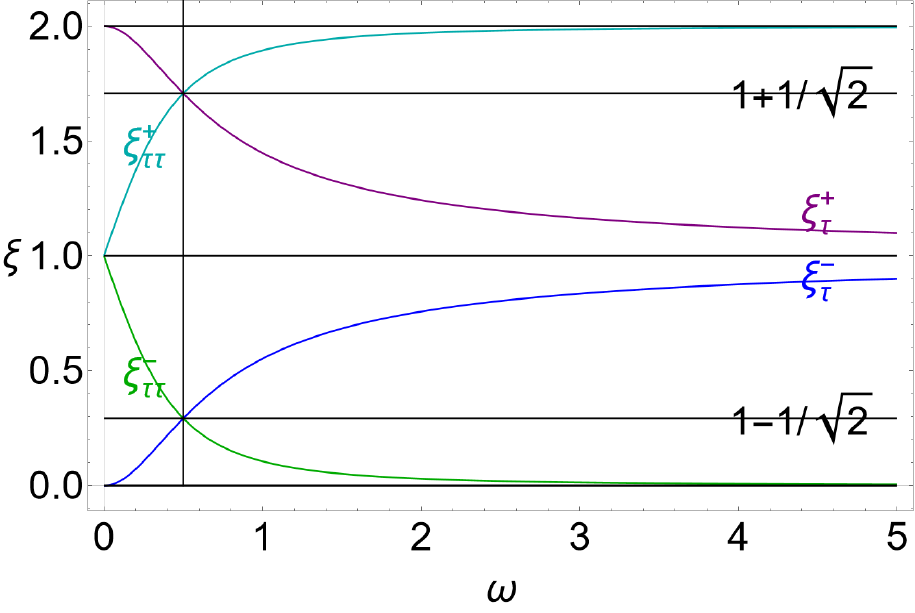}
           \caption{Black hole irreducible mass functions $M_{irr}$ (upper panel) and maximum extractable  rotational energy $\xi$ ({lower} panel) versus bundle characteristic frequency $\omega$. See the analysis of Sec.\il(\ref{Sec:sud-mirr-egi}). The functions $\xi_{\tau}^\pm$ and $\xi_{\tau\tau}^\pm$ are in Eq.\il(\ref{Eq:xtautau}), with limiting values $\xi_{\tau}^+=\xi_{\tau\tau}^+=1+1/\sqrt{2}$ and $\xi_{\tau}^-=\xi_{\tau\tau}^-=1-1/\sqrt{2}$ for $\omega=1/2$ (corresponding to the extreme Kerr \textbf{BH} ) are also shown. The functions $M_{irr}(\xi_\tau^-)$ and  $M_{irr}(\xi_{\tau\tau}^-)$ are in Eq.\il(\ref{Eq:candd}). The limiting value $M_{irr}(\xi_\tau^-)=M_{irr}(\xi_{\tau\tau}^-)=1/\sqrt{2}$ for $\omega=1/2$ (corresponding to the extreme \textbf{BH} ) is  shown. All quantities are dimensionless.}\label{Fig:PlotClipMirr}
\end{figure*}

We can   relate
  the rotational energy parameter   $\xi$ to the horizons, in the extension of the plane for extended values of $\xi$, where the following two cases are possible
\bea\nonumber
&&
\mathbf{(1)}\quad\xi^{\mp}_{\pm}=1\pm\sqrt{\frac{r_{\mp}}{2}},\quad \mbox{and}\quad \xi_\nu^{\mp}\equiv1\mp\sqrt{\frac{r}{2}},\quad\mbox{then}\quad \xi_{\tau}^{\mp}\equiv 1\mp\frac{1}{\sqrt{4 \omega ^2+1}},
\\\label{Eq:xtautau}
&&
\mathbf{(2)} \quad  \xi_\mu^{\mp}=1\mp\sqrt{1-\frac{r}{2}}, \quad \mbox{then}\quad
\xi_{\tau\tau}^{\mp}\equiv 1\mp\frac{2 \omega }{\sqrt{4 \omega ^2+1}}
\eea
--Figs\il(\ref{Fig:PlotClipMirr}). Here, $r$  is a point of the horizon curve in the extended plane.
 However, as   $\xi\propto M-M_{irr}$  the solution $\xi_-^+$ has to be considered.

 Within this parametrization, the frequency $\omega_H^+={1}/{2 \sqrt{2}}$  for the outer horizon  $r/M= 4/3$  (\textbf{BH} with  dimensionless spin $a/M={2 \sqrt{2}}/{3}$)  is also  the saddle point of $\xi_\tau^\pm$ as function of $\omega$. The maximum extractable  rotational energy decreases/increases faster with the  horizon (and \MB s) frequencies constrained  by  the limiting values $\omega_H^+={1}/{2 \sqrt{2}}$.

The  characteristic bundle (horizons)  frequencies,  can be expressed  in terms of the dimensionless parameter $\xi$ as
\bea\label{Eq:show-g}
\omega_{\pm} \equiv \frac {1  \pm \sqrt {4 (\xi -
            2)\xi (\xi - 1)^2 +
        1}} {4\sqrt {(2-\xi ) (\xi - 1)^2\xi}}.
\eea
With  the irreducible mass definitions  and Eqs\il(\ref{Eq:xtautau})
we obtain, respectively
 \bea &&\nonumber
\mathbf{(1)}\quad \frac{M_{irr}}{M}=1-\xi^{\mp}_{\pm}=\mp\sqrt{\frac{r_{\mp}}{2}},\quad\mbox{where}\quad \frac{M_{irr}(\xi_\nu^{\mp})}{M}\equiv1-\xi_\nu^{\mp}=\pm\sqrt{\frac{r}{2}}\\
&&\nonumber\mbox{and} \quad \frac{M_{irr}(\xi_{\tau}^{\mp})}{M}\equiv 1-\xi_{\tau}^{\mp}= \pm\frac{1}{\sqrt{4 \omega ^2+1}}
\\&&\label{Eq:candd}
\mathbf{(2)}\quad
\frac{M_{irr}(\xi_\mu^\mp)}{M}\equiv 1-
\xi_\mu^{\mp}\equiv\pm\sqrt{1-\frac{r}{2}},\\
&&\nonumber\mbox{and}\quad
\frac{M_{irr}(\xi_{\tau\tau}^\mp)}{M}\equiv 1-\xi_{\tau\tau}^{\mp}= \pm\frac{2 \omega }{\sqrt{4 \omega ^2+1}}
\eea
--Figs\il(\ref{Fig:PlotClipMirr}).
For $r=0$, according to Eqs\il(\ref{Eq:xtautau}), it is  $M_{irr}/M=\{1,0\}$, with  $M_{irr}/M=1/\sqrt{2}$ for $r=M$, and   $M_{irr}/M=0$ for $r=2M$.

\subsection{BH thermodynamics in the extended plane}\label{Sec:fin-BH-THEr-presenT}

Let us start from the relation  
\bea\nonumber
\frac{\delta M_{irr}}{M_{irr}}=\frac{\delta M-\delta J \omega_H^+(0)}{\sqrt{M(0)^2-\frac{J(0)^2}{M(0)^2}}},\eea
where $
\delta M_{irr}\geq 0$, thus $(\delta M-\delta J \omega_H^+(0))\geq 0$  and  all quantities are evaluated on  the outer Killing horizon for the initial \textbf{BH}.

Note, the upper bound $
\delta M_{irr}\geq 0$ sets the limit
\bea&&\nonumber
\frac{\delta J}{\delta M}\leq\frac{1}{ \omega^+_H(0)}=
\la_0(0)=
 \frac{2 r(0)}{\sqrt{(2-r(0)) r(0)}}
 \\&& (\mbox{where}\quad
\la_0(0)\geq 2,\quad  r(0)\in[1,2])
\eea
to any spin variation of the \textbf{BH} , where we  have used Eq.\il(\ref{Eq:la0-origin})
in terms of the dimensionless origin spin of the bundle and its dimensionless tangent radius  to the outer horizon curve.
It is a well known fact that limits of \textbf{BH} energy extraction
are imposed   by the \textbf{BH} horizon. Indeed,
massive  particles  or photons with  momentum $p^a$ that cross the outer horizon $r_+$
 of a Kerr \textbf{BH}  should satisfy the inequality
$-p_a  (\mathcal{L}_H^+)^a\geq 0$,
This implies that
$\mathcal{E}-\omega_H^+ \mathcal{L}\geq 0$,
for $\mathcal{E}\equiv -p_a \xi^a_{(t)}$ (energy of the particle as measured at infinity),
where $\mathcal{L} = p_a \xi^a_{(\phi)}$ is the  $\phi$ component of the particle (photon) angular momentum.
Thus, the specific angular momentum
$
 \mathcal{L}/\mathcal{E}\leq 1/\omega_H^+=\la_0$.
If the energy $\mathcal{E}$ is negative,  $\mathcal{L}$ is negative and  the  \textbf{BH} spin is reduced  (\textbf{BH} spin--down).
(The bound regulates also the super-irradiance  as  the  analogue of the Penrose process for  radiation scattering  by a
Kerr \textbf{BH}.
For a wave--mode of angular-frequency $\omega$,
$\omega$ is amplified if  $\omega\in]0,{z}/{\la_0}[$,
where  $z\in Z-\{0\}$ is  the wave angular momentum number).

More generally, the \textbf{BH}  area  is a function of the horizon radius as $A_{area}(a_{\pm})=8\pi r$\footnote{i.e.
$
 A_{area}^{\pm}\left(r_{\pm }\right)=\pm 8\pi r_{\pm }$. With  $ \delta A_{area}^- =-\delta A_{area}^+ $ or  $
 \partial_a A_{area}^{+}\delta a =-\partial_a A_{area}^{-}\delta a$.
From the relation $\delta r_ += -\delta r_ -$,  (since
 $r_ + r_ -= a^2$ and $ r_++ r_ -= 2M$) and, therefore, it holds if $\delta M=0$.}.  Considering the  variation of the horizon area (and the  irreducible masses)  for the ADM mass $M$, the area $A_{area}$ and the momentum $J$, we obtain
\bea&&\label{Eq:prop-time}
\delta M^{\mp}=\frac{\ell^{\mp}}{8\pi} \delta A_{area}^{\mp}+ \omega_H^{\mp} \delta J^{\mp},
\eea
where the following relation holds
\bea
\frac{da_{\pm}}{ dr}=\mp\left.\frac{\ell}{\omega_\pm}\right|_{a=a_\pm},
\eea
in terms of frequency and surface gravity function evaluated on the horizon curve $a_\pm$.

Explicitly, using the expressions  for the bundle frequencies and the surface gravity function,
 \bea&&\nonumber
 \omega=\frac{\sqrt{2-r}}{2 \sqrt{r}}=\frac{1}{2} \sqrt{1-4\ell},\quad\mbox{and}\quad
 \ell(r,a_\pm)=\frac{r-1}{2 r},
 \eea
 which are   related by $\ell=\omega_{\mathbf{EBH}}^2-\omega ^2$, where $\omega_{\mathbf{EBH}}=1/2$ is the extreme \textbf{BH}  horizon  frequency.

Finally, in the extended plane,  we obtain
\bea&&\label{Eq:minacc-faraoana}
\delta M=\frac{\delta A_{area} (r-1)}{2 r}+\frac{\delta J \sqrt{2-r}}{2 \sqrt{r}},\quad\mbox{and}\\&&\nonumber \delta M=\delta A_{area} \left(\frac{1}{4}-\omega ^2\right)+\delta J \omega,\quad\mbox{or}\quad  \delta M=\left(\frac{1}{4}-\frac{1}{\la_0^2}\right) \delta A_{area}+\frac{\delta J}{\la_0}
\eea
 in terms of the \MB s tangent  radius $r$,  characteristic  frequency $\omega$, and origin spin $\la_0$, where the limiting case $\la_0=2$,
$\omega=1/2$ and  $r=1$  corresponds here to the extreme Kerr \textbf{BH} , distinguishing  the relations  defined on the inner and outer horizons curves in the extended plane, where the terms relative to the surface gravity change sign.

From
   $A_{area}^-= -A_{area}^++ 4 M^2$,
we can write   $\delta A_{area}^- = -\delta A_{area}^+ + 8 M \delta M$,
(with   $\delta J=0 $ then
$	\delta M={\delta A_{area}^+  \ell_H^+}/{2 M}$). 

In the  special case of  $\delta M=0$ (invariant \textbf{BH}  mass), $\delta A^+_{area}=-\delta A_{area}^-$  and
\bea&&
\frac{\delta J}{\delta A_{area}^+} = \frac{\ell^-+\ell^+}{\omega_H^--\omega_H^+}.
\eea
Then, using  quantities as defined in the extended plane,  we obtain
\bea&& \frac{\delta J}{\delta A_{area}^{\mp}}=(\pm)\frac{\sqrt{1-a^2} }{a}=
\frac{4-\la_0^2}{4\la_0}=\frac{4 \omega^2-1}{4 \omega}=\frac{1-r}{\sqrt{r(2-r)}},
\eea
which represents the variation of the angular momentum versus area in terms of the initial bundle tangent point of the horizon $(a,r)$,
or, alternately, the characteristic frequency $\omega$ or its origin spin $\la_0$ in dependence from the poloidal angle $\sigma$. Note that the limiting case $\la_0=2$ (a \textbf{NS}) is equivalent to
$\omega=1/2$ for the extreme Kerr spacetime
i.e. $(a=1, r=1)$.

\section{Final remarks}\label{Sec:final-remarks}
Metric bundles relate all the geometries  and all the orbits with  equal photon  circular orbital frequency.  These geometries always include  \textbf{BHs} and in some cases also \textbf{NS}s.
In the  Kerr background, for example, the bundles  characteristic frequencies coincide with the frequencies of the Kerr Killing horizons.
In this way, \textbf{NS} solutions are related to  \textbf{BH}  solutions (in the bundle). So, the \textbf{BH}  horizon distinguishing the characteristic bundle frequency and some properties typical of  the \textbf{NS} light surfaces can be explained in terms of \MB s, and their relation to the Killing horizons are expressed in the extended plane.

  \MB s    have a natural application in  \textbf{BH} astrophysics since they  are  constructed by photon (circular) orbits    which can be measured by  observers\footnote{Furthermore, the frequencies of stationary observers, bounded by the light surfaces defining  the bundles, determine many aspects of   \textbf{BH} accretion  configurations and jets launching and  collimation, Blandford-Znajek process (a further mechanism for the extraction of rotational energy from a spinning \textbf{BH} , where  there is a magnetized accretion disk  and a base engine powering  jet launching around super-massive Kerr \textbf{BH}s), accretion disks or  the  Grad-Shafranov equation for the force free magnetosphere around \textbf{BHs}.}.
  Some aspects of the background   causal structure are determined by the crossings of metric bundles in the extended plane,  which has been proved to  be  essentially determined by the horizon curves.
Replicas connect  different spacetime regions allowing to explore, for example, regions close to the  \textbf{BH} horizon.

Horizon replicas are  bundle orbits,   other  than that of \textbf{BH} horizons,  but characterized by  photons  with orbital frequency  equal to the (inner or outer) horizon frequency  of the \textbf{BH} spacetime.
From an observational viewpoint, it is worth to note that \MB s can be  used also to characterize the geometry and causal structure in the regions close to the \textbf{BH} poles and rotational axis--\cite{nuclear,GRG-letter}.
In fact, this constitutes an aspect of information extraction of the \textbf{BH} properties into the  region accessible to  far away observers.
Horizon replicas depend on the angle $(\sigma\in[0,1])$ with respect to  the rotational axis, and the exploration of this region  may have important implications for the knowledge of spacetimes structures closed to the singularity.

Replicas can also be defined for counter-rotating photons, i.e., negative frequencies with respect to the positive frequencies in the positive section of the extended plane.
Extended  planes  and metric bundles allow us to  connect different points of  one geometry,  but also different  geometries. Through the notion of replica we highlight those properties of a \textbf{BH} horizon, which can be replicated  in other points of the same or different spacetimes, providing  a new  and global frame for the  interpretation of these metrics and, in particular, of \textbf{NS} solutions  which result connected, through \MB s curves tangency properties,  to the \textbf{BH}  horizons in the extended plane. This  property establishes  a \textbf{BHs}--\textbf{NSs} connection,  highlighting    important properties  of the Kerr geometries.
 The extended plane is, therefore,  equivalent to a function relating  the  characteristic bundle (\textbf{BH}s  horizon frequencies) to  the
(bundles origin) spin leading to an alternative definition  of the Killing  horizons\footnote{Replica are studied with
the analysis of self-intersections of the bundles curves in the extended plane, in the same geometry (horizon confinement)  or
intersection of bundles curves  in different geometries.
}.
The bottlenecks and remnants (or  pre-horizons regime)  can also be interpreted  in terms of metric bundles,   describing  some properties of the Killing horizons in  axially symmetric spacetimes and event horizons in the spherically symmetric case.

We have shown that it is possible to write     aspects  of   classic black hole thermodynamics reformulated  in terms of light surfaces
and  horizons,  relating the initial and final states of a \textbf{BH} transition, expressed  by the laws of \textbf{BH}  thermodynamics in terms of the light surfaces.

The extended plane   can represent  a significant global frame   also for the analysis of   \textbf{BH}s   transitions, where the
 \textbf{MB}s utility  lies in enlightening   spacetime properties    emerging in the extended plane,  related to the local causal structure and \textbf{BH}  thermodynamics. A (stationary) \textbf{BH}  transition from a state to another is defined  by a transition  of its
characteristic parameters,  and it is  regulated by the  laws of \textbf{BH}  thermodynamics, governed  by  the values   \textbf{BH} state  horizon frequency and surface gravity, before the  transition.
We can express    the    \textbf{BH}  surface gravity   using the  light surfaces of the corresponding  geometry for  a \textbf{BH}
in equilibrium, i.e.,{ stationary\footnote{In fact, while surface gravity for stationary \textbf{BH}s   is well defined (as there is a well defined Killing horizon, where the Killing vector is normalized to  unit  at spatial infinity),  this is not the case for dynamical \textbf{BH}s \cite{NAY1,PMK1}.}.}
Initial and final  states of \textbf{BH} transitions can be related using \MB s,  or rephrased  alternatively, the  \textbf{BH} states are given   in terms of the bundle characteristic frequency.
The analysis through bundles enlighten the possible existence of  privileged or not allowed state transitions.

%


\begin{thebibliography}{99}

\bibitem{BCH}
  J. M. Bardeen, B. Carter,  S. W. Hawking, 
 Commun. Math. Phys.  \textbf{31},  2 (1973).


\bibitem{Bekenstein73}J. D. Bekenstein, Phys. Rev. D \textbf{7}, 2333 (1973).
\bibitem{Bekenstein75}J. D. Bekenstein, Phys. Rev. D \textbf{12}, 3077 (1975).
\bibitem{BaPRL}
S. Bhattacharya, A. Lahiri,
Phys. Rev. Lett. \textbf{99}, (20) 201101  (2007).

\bibitem{Birkhoff}G. D.  Birkhoff, Relativity and Modern Physics,  Cambridge, Massachusetts: Harvard University Press. LCCN 23008297  (1923).
\bibitem{Puls}
O. Brodbeck, M. Heusler,  N. Straumann,
 Phys. Rev. D \textbf{53}, 754--761, (1996). 

\bibitem{Puls1}O. Brodbeck,  M. Heusler, N. Straumann,  M. Volkov,
Phys. Rev. Lett. \textbf{79}, 4310--4313, (1997).

\bibitem{Carterspinning}
 B. Carter,  
   Phys. Rev. Lett. \textbf{26} (6), 331--333 (1971).

\bibitem{Chakraborty:2016mhx}
   C. Chakraborty, M. Patil, et al 
   Phys.\ Rev.\ D {\bf 95}, 8,  084024 (2017).
  \bibitem
{Ruffini}
 D. M. Christodoulou\&  R. Ruffini, Phys. Rev. D \textbf{4}, 3552 (1971).



  \bibitem{DafermosLuk}
M. Dafermos \& J. Luk,
[arXiv:1710.01722 [gr-qc]]).

\bibitem{Daly0}
R. A. Daly, Ap.J. \textbf{691}, L72-L76,  1 (2009).

\bibitem{Daly2}
R. A. Daly, Mont. Notice R. astr. Soc.  \textbf{414}, 1253-1262 (2011).

\bibitem{Daly3}
R. A. Daly \&T. B. Sprinkle, Mont. Notice R. astr. Soc.  \textbf{438}, 3233-3242 (2014).




\bibitem{de-Felice1-frirdtforstati}
F. de Felice,
Mont. Notice R. astr. Soc.  \textbf{252}   197-202 (1991).

\bibitem{de-Felice4-overspinning} F.  de Felice,   L.  Sigalotti, Ap.J. \textbf{389}, 386-391 (1992)

\bibitem{de-FeliceKerr}  F. de Felice, Class. Quantum Grav. \textbf{11},  1283-1292 (1994).


  \bibitem{de-Felice-first-Kerr}
  F. de Felice and S. Usseglio-Tomasset, Class. Quantum Grav. \textbf{8}., 1871-1880 (1991).


\bibitem{de-Felice3} F. de Felice,  S. Usseglio-Tomasset, Gen. Rel. Grav. \textbf{24},  10 (1992).


\bibitem{de-Felice-anceKerr}F.  de Felice, S. Usseglio-Tomasset, Gen. Rel. Grav. \textbf{28},   2 (1996).


\bibitem{de-Felice-mass}  F. de Felice,   Y. Yunqiang, Class. Quantm Grav. \textbf{10}, 353-364 (1993).





 \bibitem{FRW}H. Friedrich, I. Racz, R. Wald, 
 Commun. Math. Phys. \textbf{204}, 691--707 (1999).

\bibitem{GarofaloEvans}
D. Garofalo, D. A. Evans, R. M. Sambruna,
Mont. Notice R. astr. Soc. \textbf{406}, 975-986 (2010).

\bibitem{soft3}
 S. Haco, S. W.  Hawking, M. J. Perry,  A. Strominger,  
  JHEP \textbf{(12)}: 98 (2018).
\bibitem{Hawking71}S. W. Hawking, Phys. Rev. Lett. \textbf{26}, 1344 (1971).
\bibitem{H72152}S. W.  Hawking, Commun. Math. Phys. \textbf{25}, 152  (1972).
\bibitem{Hawking74}S. W. Hawking, Nature \textbf{248}, 30-31, (1974).
\bibitem{Hawking75}S. W. Hawking, Comm. Math. Phys. \textbf{43}, 199 (1975) Erratum - ibidem \textbf{46}, 206 (1976).


\bibitem{H77}S. W. Hawking, 
Scientific American \textbf{236},  1, 34--40 (1977).
\bibitem{Hawking05}S. W.  Hawking,   
 Phys. Rev. D. \textbf{72} (8), 4 (2005).
\bibitem{HE}S. W. Hawking, G. F. R. Ellis, \emph{The large scale structure of space--time}, Cambridge
University Press, Cambridge, (1973).



\bibitem{soft1}S. W. Hawking, M. J. Perry, A. Strominger,  
 Phys. Rev. Lett. \textbf{116} (23), 231301  (2016).




  \bibitem{soft2}
G. T.  Horowitz,  
 Physics \textbf{9}. (2016).






\bibitem{Isi:2020tac}
M.~Isi, W.~M.~Farr, M.~Giesler, et al.
Phys. Rev. Lett. \textbf{127}, 1, 011103  (2021).
\bibitem{Israel1}
W.  Israel, 
  Phys. Rev. \textbf{164} (5), 1776--1779  (1967).
 \bibitem{Israel2}
W.  Israel,   
  Commun. Math. Phys. \textbf{8} (3), 245--260 (1968).
  \bibitem{J-S09}  T.  Jacobson\& T. P. Sotiriou, Phys. Rev. Lett. \textbf{103}, 141101
(2009).
\bibitem{Jacobson:2010iu}
  T.~Jacobson \& T.~P.~Sotiriou,
  J.\ Phys.\ Conf.\ Ser.\  {\bf 222},  012041 (2010).



\bibitem{Kunz2} B. Kleihaus and J. Kunz,
Phys. Rev. Lett.
\textbf{79}, 1595-1598, (1997).

\bibitem{malament} D. B. Malament, J. Math. Phys. \textbf{18}, 1399 (1977).


\bibitem{Mars}
M. Mars et al.,  Class. Quantum Grav. \textbf{35}, 155015, (2018).
\bibitem{Mukherjee:2018cbu}
   S.  Mukherjee,   R.~K. Nayak,
  Astrophys.\ Space Sci.,\  {\bf 363}, 8  163 (2018).

\bibitem{NAY1} A. Y. Nielsen,  
Class. Quantum Grav.  \textbf{25} (8), 085010 (2008).

\bibitem{Penrose69} R. Penrose, Riv. Nuovo Cim. \textbf{1}, 252-276 (1969).



\bibitem{PMK1} M. Pielahn, G. Kunstatter, A. B. Nielsen,   
Phys. Rev. D. \textbf{84} (10), 104008(11) (2011).

\bibitem{JP16}J. Polchinski,
Contribution to: TASI, 353-397  (2015).

\bibitem{LQG}D. Pugliese, G. Montani, Entropy \textbf{22}(4), 402 (2020).







\bibitem{observers}
D. Pugliese, H. Quevedo
  Eur.\ Phys.\ J. {C}, { \textbf{78}}  1 69 (2018).

  \bibitem{remnants}
D. Pugliese, H. Quevedo,
Eur. Phys. J. {C},  \textbf{79} 3 209 (2019).

  \bibitem{nuclear}
D.~Pugliese, H.~Quevedo,
Nucl. Phys. B \textbf{972}, 115544 (2021).

\bibitem{GRG-letter}
D.~Pugliese, H.~Quevedo,
Gen. Rel. Grav. \textbf{53}, 10, 89 (2021).

\bibitem
{bundle-EPJC-complete}
D. Pugliese, H. Quevedo,
 Eur. Phys. J. {C} \textbf{81}  3 258 (2021).


\bibitem{wormhole}
D. Pugliese, H. Quevedo, 
  Eur. Phys. J. {C} \textbf{82}, 1090 (2022).





\bibitem{ella-correlation}D. Pugliese, Z. Stuchl\'\i{}k,
Class. Quant. Grav., \textbf{38}  14  14  (2021).








\bibitem{Wi2}
 S. A. Ridgway,  E. J. Weinberg, 
Phys. Rev. \textbf{D}, 52, 3440-3456, (1995).


\bibitem{Riegert}
R. J. Riegert,
Phys. Rev. Lett. \textbf{53}, 315 (1984).

\bibitem{Robinson}D. C. Robinson, Gen. Rel. Grav. \textbf{8}, 695--698  (1977).



\bibitem{Simpenrose}
M. Simpson\& R. Penrose, Int. J.
Theor. Phys. \textbf{7} (1973).

\bibitem{Smarr} L. Smarr, Phys. Rev. Lett. \textbf{30}, 71 (1973). Erratum: [Phys. Rev. Lett. 30,
521 (1973)].


 \bibitem{SV96}A. Strominger, C. Vafa, 
 Phys. Lett. B. \textbf{379} (1--4), 99--104  (1996).


\bibitem{LS}L. Susskind, \emph{The Black Hole War: My Battle with Stephen Hawking to Make the World Safe for Quantum Mechanics},
Little, Brown, (2008).



\bibitem{Tanatarov:2016mcs}
  I.~V.  Tanatarov and O.~B. Zaslavskii,
  Gen.\ Rel.\ Grav.\  {\bf 49}, 9,  119 (2017).
Wald:1999xu\bibitem{Wald:1999xu}
  R.~M.~Wald,
  Class.\ Quant.\ Grav.\  {\bf 16}, A177  (1999).


\bibitem{WW}R. M. Wald,
Living Rev. Relativ.  \textbf{4}(1), 6 (2001).




    \bibitem{Zaslavskii:2018kix}
  O.~B.~Zaslavskii,
  Phys.\ Rev.\ D {\bf 98}, 10,  104030  (2018).


 \end{thebibliography}
\end{document}